\documentclass[twocolumn]{aastex62}

\graphicspath{{./}{figures/}}

\usepackage{dirtytalk}
\usepackage{amsmath}
\usepackage{graphics}
\usepackage{rotating}
\usepackage[]{hyperref}
\usepackage{cleveref}
\crefformat{section}{\S#2#1#3} 
\crefformat{subsection}{\S#2#1#3}
\crefformat{subsubsection}{\S#2#1#3}
\usepackage{multirow}
\usepackage{enumitem}

\accepted{18 December 2022}

\shorttitle{Multiple shock fronts in RBS~797}
\shortauthors{Ubertosi et al.}

\begin{document}

\title{Multiple shock fronts in RBS~797: the \textit{Chandra} window on shock heating in galaxy clusters}

\correspondingauthor{F. Ubertosi}
\email{francesco.ubertosi2@unibo.it}

\author[0000-0001-5338-4472]{F. Ubertosi}
\affil{Dipartimento di Fisica e Astronomia, Università di Bologna, via Gobetti 93/2, I-40129 Bologna, Italy}
\affil{INAF, Osservatorio di Astrofisica e Scienza dello Spazio, via P. Gobetti 93/3, 40129 Bologna, Italy}

\author[0000-0002-0843-3009]{M. Gitti}
\affil{Dipartimento di Fisica e Astronomia, Università di Bologna, via Gobetti 93/2, I-40129 Bologna, Italy}
\affil{Istituto Nazionale di Astrofisica (INAF) - Istituto di Radioastronomia, via Gobetti 101, I-40129 Bologna, Italy}

\author[0000-0001-9807-8479]{F. Brighenti}
\affil{Dipartimento di Fisica e Astronomia, Università di Bologna, via Gobetti 93/2, I-40129 Bologna, Italy}
\affil{University of California Observatories/Lick Observatory, Department of Astronomy and Astrophysics, Santa Cruz, CA 95064, USA}

\author[0000-0001-5226-8349]{M. McDonald}
\affil{Kavli Institute for Astrophysics and Space Research, Massachusetts Institute of Technology, Cambridge, MA 02139, USA}

\author[0000-0003-0297-4493]{P. Nulsen}
\affil{Chandra X-ray centre, Smithsonian Astrophysical Observatory, 60 Garden Street, Cambridge, MA 02143, USA}
\affil{ICRAR, University of Western Australia, 35 Stirling Hwy, Crawley, 
WA 6009, Australia}

\author[0000-0002-2808-0853]{M. Donahue}
\affil{Department of Physics and Astronomy, Michigan State University, East Lansing, MI 48824, USA}

\author[0000-0003-4195-8613]{G. Brunetti}
\affil{Istituto Nazionale di Astrofisica (INAF) - Istituto di Radioastronomia, via Gobetti 101, I-40129 Bologna, Italy}

\author[0000-0002-3984-4337]{S. Randall}
\affil{Center for Astrophysics, Harvard \& Smithsonian, 60 Garden St., Cambridge, MA 02138, USA}

\author[0000-0003-2754-9258]{M. Gaspari}
\affiliation{Department of Astrophysical Sciences, Princeton University, Princeton, NJ 08544, USA}

\author[0000-0003-4117-8617]{S. Ettori}
\affil{INAF, Osservatorio di Astrofisica e Scienza dello Spazio, via P. Gobetti 93/3, 40129 Bologna, Italy}
\affil{INFN, Sezione di Bologna, viale Berti Pichat 6/2, 40127 Bologna, Italy}

\author[0000-0002-2238-2105]{M. Calzadilla}
\affil{Kavli Institute for Astrophysics and Space Research, Massachusetts Institute of Technology, Cambridge, MA 02139, USA}

\author[0000-0003-1581-0092]{A. Ignesti}
\affil{INAF - Astronomical Observatory of Padova, vicolo dell'Osservatorio 5, IT-35122 Padova, Italy}

\author[0000-0003-0312-6285]{L. Feretti}
\affil{Istituto Nazionale di Astrofisica (INAF) - Istituto di Radioastronomia, via Gobetti 101, I-40129 Bologna, Italy}

\author{E. L. Blanton}
\affil{Institute for Astrophysical Research and the Department of Astronomy, Boston University, Boston, MA 02215, USA}




\begin{abstract}
Using $\sim$427 ks of \textit{Chandra} observations, we present a study of shock heating and ICM cooling in the galaxy cluster RBS 797. We discover three nested pairs of weak shocks at roughly  50 kpc, 80 kpc and 130 kpc from the center. The total energy associated with the shocks is $\sim6\times10^{61}$ erg, with the central AGN driving a pair of weak shocks every 20-30 Myr with a power $P_{\text{sh}}\approx10^{46}$ erg s$^{-1}$. Based on its morphology and age ($\sim$30 Myr), the inner cocoon shock is associated with the four equidistant X-ray cavities previously discovered. From the thermodynamic analysis of the inner 30 kpc, we find evidence for ICM condensation into colder gas between and behind the X-ray cavities. The total AGN mechanical power (cavities and shocks) of $3.4\times10^{46}$ erg s$^{-1}$ can balance the ICM radiative losses, estimated as $L_{\text{cool}} = 2.3\times10^{45}$ erg s$^{-1}$. By building plots of $P_{\text{cav}}\,\text{vs.}\,L_{\text{cool}}$, $P_{\text{shock}} \,\text{vs.}\, L_{\text{cool}}$ and $P_{\text{tot}} \,\text{vs.}\, L_{\text{cool}}$ for RBS~797 and 14 other galaxy clusters, groups and elliptical galaxies where both cavities and shocks are detected, we verify that the most powerful outbursts are found in the strongest cooling systems. Ultimately, we observe that the mechanical power of the AGN exceeds the gas radiative losses by a factor that is different for FR~I and FR~II radio galaxies, being less than a few tens for FR~Is (as RBS 797) and more than roughly a hundred for FR~IIs.
\end{abstract}


\section{Introduction} \label{sec:intro}
In the last two decades, multi-wavelength observations and theoretical studies of galaxy clusters, galaxy groups and elliptical galaxies have revolutionized the understanding of the effects of Active Galactic Nuclei (AGNs) on the surrounding gaseous halos (e.g., for reviews \citealt{2003ARA&A..41..191M,2004cgpc.symp..143D,2007ARA&A..45..117M,2012ARA&A..50..455F,2012AdAst2012E...6G,2012NJPh...14e5023M,2020NatAs...4...10G,2021Univ....7..142E,2022PhR...973....1D}). In particular, it has been determined that a feedback loop regulates the interaction between the intracluster medium (ICM) and the central AGN in the brightest cluster galaxy (BCG). According to this scenario, the radiative losses of the ICM lead to the formation of dense central regions of low temperature gas (the so called \textit{cool cores}; see e.g. \citealt{2010A&A...513A..37H}), that may constitute large reservoirs of gas to fuel both star formation and the central supermassive black holes (SMBH). However, in cool cores the amount of cold material actually observed to be accumulating around BCGs is less than what is expected from uninterrupted cooling of the X-ray emitting atmosphere \citep{1993MNRAS.264L..25B,2004cgpc.symp..143D,2006PhR...427....1P}. This discrepancy can be accounted for by considering that AGN activity is triggered in response to the build-up of these reservoirs, and is able to regulate the thermodynamic state of the gas (e.g., \citealt{2007ARA&A..45..117M,2012NJPh...14e5023M,2016ApJ...830...79M,2018ApJ...858...45M,2022PhR...973....1D}). 
\\ \indent On the one hand, the relativistic AGN jets have a tremendous impact on the gas morphology and thermodynamics, carving large holes (the X-ray cavities) and driving shocks in the ICM, which have been observed in the X-ray band (e.g., \citealt{2000A&A...356..788C,2007ApJ...659.1153W,2011ApJ...732...13G,2015ApJ...805..112R,2014MNRAS.442.3192V,2018ApJ...855...71S}). These outbursts typically inject between $10^{56} - 10^{62}$ erg in the environment, with the more energetic events being found on average in the largest cool cores \citep{2004ApJ...607..800B,2006ApJ...652..216R,2018ApJ...858...45M}. Such an outcome of AGN feedback may explain how further cooling of the ICM can be limited. 
\\ \indent On the other hand, AGN activity can also stimulate cooling of the ICM. The mechanical thrust imparted to the gas by the rising X-ray cavities can uplift the central cool material to several tens of kpc, where cooling times and dynamical times become comparable (e.g., \citealt{2008A&A...477L..33R,2015ApJ...802..118B,2016ApJ...830...79M}). Alternatively, the turbulence injected by the nuclear activity may trigger local compressions of the gas (e.g., \citealt{2002ApJ...573..542B,2011MNRAS.411..349G,2012MNRAS.424..190G}). In such cases, as thermal instabilities may ensue, further cooling of the ICM can be stimulated, creating warm or cold filaments (e.g., \citealt{2010ApJ...721.1262M,2019MNRAS.490.3025R,2022ApJ...928..150T,2022arXiv220107838O,2022ApJ...940..140C}). 
\\ \indent Regarding the observational footprints of feedback, AGN-inflated cavities are the most evident outcomes of radio lobe expansion in the ICM, and have been observed in numerous galaxy clusters, groups and elliptical galaxies (e.g., \citealt{2004ApJ...607..800B,2006ApJ...652..216R,2016ApJS..227...31S}). Since the enthalpy of the bubbles is a proxy for the energy deposited by
the jet, the study of large samples of radio galaxies in clusters has provided constraints on the relation between the synchrotron radio power and the kinetic power (e.g., \citealt{2004ApJ...607..800B,2008ApJ...686..859B,2012MNRAS.427.3468B,2010ApJ...720.1066C,2011ApJ...735...11O}). Specifically, from these relations it has been determined that for both high radio luminosity radio galaxies (or Fanaroff-Riley Type IIs, FR~IIs) and low radio luminosity radio galaxies (Fanaroff-Riley Type Is, FR~Is, \citealt{1974MNRAS.167P..31F}), the total radio power roughly scales with the mechanical luminosity, which is usually larger by a factor of 100 (see also \citealt{2012AdAst2012E...6G,2016PASJ...68...26F}), implying that the mechanical feedback response largely exceeds the radiative one. Thus, the investigation of the hot gas surrounding radio galaxies in clusters and groups can be undertaken to probe feedback, SMBH jet formation, and AGN mechanical power.
\\ \indent In a fraction of clusters, groups, and elliptical galaxies, multiple X-ray cavities at different distances from the center have been detected, indicating successive episodes of radio activity in the BCG (e.g., \citealt{2005MNRAS.363..891F,2005MNRAS.364.1343D,2007ApJ...659.1153W,2014MNRAS.442.3192V,2015ApJ...805...35H,2015ApJ...805..112R}). Such discoveries made it possible both to probe AGN response to ICM cooling over time, and to recover valuable information on the duty cycle of AGNs. In particular, it has been determined that on average, every few tens of Myr the central AGN initiates a new cycle of activity, inflating new lobes and excavating additional pairs of cavities (e.g., \citealt{2004ApJ...607..800B,2012MNRAS.427.3468B,2013ApJ...768...11B,2014MNRAS.442.3192V,2015ApJ...805..112R,Biava2021}). While several aspects related to X-ray cavities are yet to be fully understood (see e.g., \citealt{2021Univ....7..142E,2022PhR...973....1D}), their general role in heating the environment has been extensively investigated and verified.
\\ \indent More uncertain is the role of shock heating in galaxy clusters. Simulations of jet expansion in the ICM indicate that besides excavating the bubbles, the gas-piercing jet also drives weak shocks (with Mach numbers between 1.1 - 1.3) that typically assume the shape of cocoons surrounding the X-ray cavities (e.g., \citealt{2005ApJ...634L.141H,2006ApJ...643..120B,2007MNRAS.380L..67B,2018A&A...617A..58C,2019MNRAS.483.2465M,2020MNRAS.498.4983W}). The energy injected in shocks is deposited not only along the jet axis (or ahead of the X-ray cavities), but is rather distributed over the whole azimuth (e.g., \citealt{2006ApJ...638..659M,2006ApJ...643..120B,2019MNRAS.483.2465M,2022arXiv220606402H}). Besides the power required to inflate X-ray cavities, considering also the mechanical power of weak shocks provides an opportunity to truly test the efficiency of AGNs in effectively keeping the ambient gas from rapid cooling (e.g., \citealt{2012NJPh...14e5023M,2015ApJ...805..112R,2021Univ....7..142E}). However, direct observational evidence of weak shocks in galaxy clusters is rare. Weak shocks driven by central AGNs have been discovered in roughly a dozen systems \citep{2019MNRAS.484.3376L}, all targeted with deep \textit{Chandra} exposures, which is lower than the number of known X-ray cavities by more than an order of magnitude (e.g., \citealt{2016ApJS..227...31S}). Given the paucity of detected shocks, little is known about their role in the feedback cycle. By analyzing a collection of 13 objects with detected shocks and X-ray cavities, \citet{2019MNRAS.484.3376L} found that the AGN mechanical energy is roughly equally divided between shocks and X-ray cavities. Nevertheless, it would also be interesting to dissect the role of shock heating and energy partition with cavities over time, which would allow the stability of the jet kinetic power across succeeding episodes of AGN activity to be addressed more robustly. So far, the unique case of the galaxy group NGC~5813 \citep{2011ApJ...726...86R,2015ApJ...805..112R}, where three groups of concentric and aligned shock fronts, each associated with expanding X-ray cavities, has demonstrated that the average jet mechanical power can be stable over roughly 50 Myr. Additionally, \citet{2015ApJ...805..112R} found that shocks alone can compensate for the radiative losses of the ICM, indicating that shock heating may be a relatively important contribution to the AGN/ICM feedback cycle. While these results indicate that the intra-group medium of galaxy groups is strongly affected by AGN feedback, these systems also have shallower gravitational potentials compared to galaxy clusters (e.g., \citealt{2010ApJ...714..218G,2010MNRAS.406..822M}), thus the effect of AGN-launched shocks on the gas may be more pronounced. As such, the conclusions based on observations of groups may not be  applicable to galaxy clusters. 
The detection of \textit{multiple} shock fronts driven by AGN activity in a galaxy cluster would expand the number of objects with known weak shocks but more importantly, would allow us to address the question of how much heating by weak shocks contributes to the feedback cycle in clusters of galaxies.
\subsection{The galaxy cluster RBS~797}\label{ourtarget}
An interesting cool core galaxy cluster to investigate these topics is RBS~797 ($z=0.354$), located at R.A. 09:47:12.76, decl. +76:23:13.74. The cluster hosts a FR~I radio galaxy at its center, showing multiple radio lobe pairs that are misaligned by roughly 90$^{\circ}$ (see \citealt{2006A&A...448..853G,2011ApJ...732...71C,2013A&A...557L..14G}), with the largest and brightest lobes extended in the east (E) - west (W) direction, smaller lobes in the north (N) - south (S) direction and perpendicular jet pairs in the inner $\sim$10 kpc. \citet{2013A&A...557L..14G} proposed that the multifaceted morphology of the radio galaxy is caused either by rapid reorientation of the AGN jets between different cycles of central radio activity, or by the presence of twin active SMBHs in the core of the BCG. Early \textit{Chandra} observations revealed that deep E-W X-ray cavities are associated with the E-W radio lobes \citep{2001A&A...376L..27S,2006A&A...448..853G,2012ApJ...753...47D}, and that the large mechanical power of the bubbles (\citealt{2011ApJ...732...71C} estimated $\approx3-6\times10^{45}$ erg s$^{-1}$) is of the order of the radiative losses in the ICM, indicating that feedback may be efficient in this galaxy cluster. Moreover, \citet{2011ApJ...732...71C} noted the presence of two surface brightness edges in the ICM. The first one surrounds the E-W X-ray cavities at $\sim$50 kpc from the center, and has been proposed to be a combination of cool gas rims encasing the X-ray cavities, and of a cocoon shock surrounding the bubbles. However, due to the insufficient number of counts collected by the old \textit{Chandra} exposures, no detailed morphological or spectral study of this putative shock has been performed. The second edge was tentatively identified at $\sim$80 kpc from the center, but again no classification was possible. Overall, the above findings point to RBS~797 being an interesting target to study the history of AGN activity in galaxy clusters and the effect of shock heating on the ICM throughout the successive central radio activities. 
\\ \indent Recently, we presented the first results from the analysis of the deeper \textit{Chandra} observations (Cycle 21 LP proposal, 420 ks, PI: Gitti, plus the previous $\sim$50 ks observations), focused on the cavity system only \citep{2021ApJ...923L..25U}. We found that the N-S radio lobes have also inflated X-ray cavities at the same projected distances as the deep E-W ones. The geometry of the inner $\sim$50 kpc of RBS~797 is thus peculiar, showing equidistant, centrally symmetric and perpendicular X-ray cavities. By measuring the age of the X-ray cavities, we found that the two outbursts are nearly coeval, with a time difference of $\lessapprox10$ Myr, which is consistent with both scenarios of a rapid reorientation (see also \citealt{2021arXiv211111906S}) and a coeval activity of binary AGNs. 
\linebreak
\\ \indent In this article we present the full analysis of the old and new \textit{Chandra} observations of RBS~797, focusing on characterizing the whole cool core region and investigating AGN feedback by both shocks and cavities. This article is organized as follows: Section \ref{sec:data} describes the observations used in this work. Section \ref{sec:result} presents the search and characterization of weak shocks propagating in the ICM. Section \ref{ssec:genicm} presents the X-ray analysis of the cluster, with specific subsections focused on: the investigation of radial spectral profiles (\S\ref{subsec:radia}), maps of thermodynamic quantities (\S\ref{subsec:spect}), the ICM abundance distribution (\S\ref{subsect:metal}). In Section \ref{subsec:point} we determine the spectral properties of the central X-ray point source. In Section \ref{sec:disc} we discuss our results, considering the implications of our findings on the history of AGN activity (\S\ref{subsec:epfeed}), on the heating and cooling balance (\S\ref{subsec:heatcool}), and on the role of shock heating in galaxy clusters, groups and elliptical galaxies (\S\ref{subsec:sampleshock}). Finally, we summarize our conclusions in Section \ref{sec:conc}. \\ \indent We assume a $\Lambda$CDM cosmology with H$_{0}$=70 km s$^{-1}$ Mpc$^{-1}$, $\Omega_{\text{m}}$=0.3 and $\Omega_{\Lambda}$=0.7, which gives a scale of 4.9 kpc arcsec$^{-1}$ at z=0.354. Uncertainties are reported at $1\sigma$, unless otherwise stated. Position angles (P.A.) of ellipses are defined northward from west along the major axis. The radio spectral index $\alpha$ is defined as $S_{\nu}\propto\nu^{-\alpha}$.

\section{The Data} \label{sec:data}

\subsection{X-ray - Chandra} \label{subsec:xray}
\begin{figure*}[ht]
	\centering
	\gridline{\fig{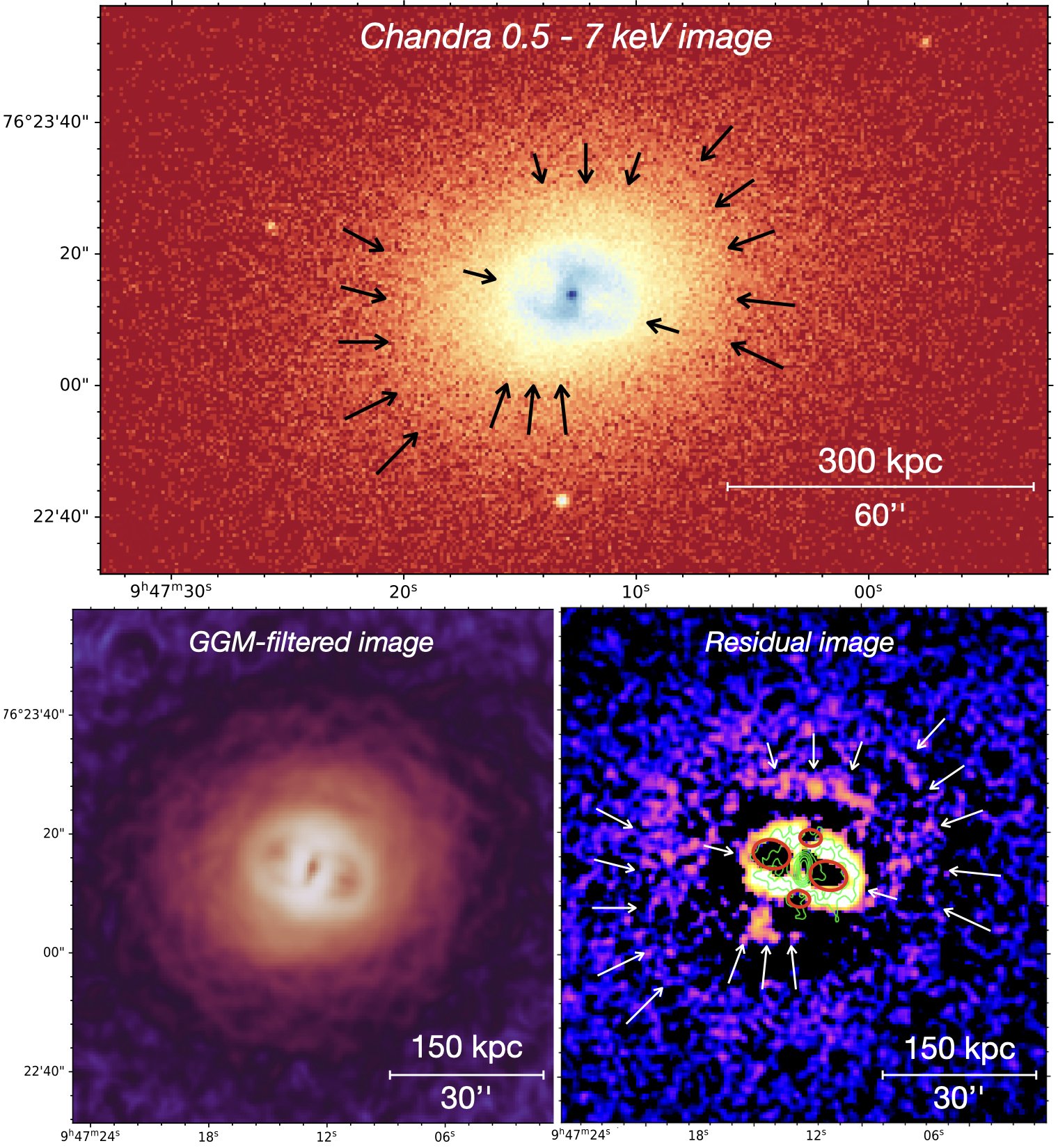}{0.98\linewidth}{}}
	\caption{\textit{Top panel:} Background subtracted, exposure corrected \textit{Chandra} image of RBS~797 in the 0.5 - 7 keV band. \textit{Bottom left panel:} \textit{Chandra} image filtered with the GGM method using $\sigma=2''$. \textit{Bottom right panel:} Double $\beta$-model residual image of RBS~797, highlighting the four X-ray cavities (red ellipses). The green contours at 1.4 GHz (at 3, 5, 10, 20, 40, 75, 150 $\times$ the rms of 0.02 mJy/beam, at $\sim$1.5$''$ resolution; see \citealt{2006A&A...448..853G}) show the morphology of the central radio galaxy. In the top and bottom right panels, the arrows indicate the position of the surface brightness edges at progressively increasing distances from the center. See \S\ref{sec:result} for details.}
	\label{fig:fig1}
\end{figure*}

\noindent The data have been reprocessed using CIAO-4.13 and CALDB-4.9.6. The available 15 ObsIDs (summarized in Tab. \ref{OBSIDS}) sum up for a total, uncleaned exposure time of 458 ks. In our region of interest (within $\sim$500 kpc from the center), the available 0.5 - 7 keV total exposure contains $\sim$300,000 net counts, which allow for a thorough analysis of the cool core region. The removal of background flares reduced the total exposure by $\sim$9\% to roughly 427 ks. To correct the astrometry of the 15 ObsIDs, the longest observation (ObsID 22932) was shifted (using the \texttt{wcs$\_$match} tool) so that the coordinates of the central X-ray point source match those of the AGN from high resolution radio observations (RA: 09 47 12.76, DEC: +76 23 13.74, \citealt{2013A&A...557L..14G}). Then, the other ObsIDs were reprojected to match the longest one. Background files were obtained from blank sky event files, normalized to the 9-12 keV count rate of the observations. 
More details on the \textit{Chandra} data reduction procedures can be found in \citet{2021ApJ...923L..25U}.
\\ \indent We produced a merged, exposure-corrected, background subtracted \textit{Chandra} image in the 0.5-7 keV band with the \texttt{merge$\_$obs} script, that reprojects and combines multiple ObsIDs. The resulting image is shown in Fig. \ref{fig:fig1}. Notable features in the image have been investigated using \texttt{CIAO} and \texttt{Proffit}, while spectral fitting (in the 0.5 - 7 keV band, with a binning of 25 counts per bin) has been performed using \texttt{Xspec12.10}, selecting the table of solar abundances of \citet{2009ARA&A..47..481A}. An absorption model (\texttt{tbabs}) was always included to account for Galactic absorption, with the column density fixed at $N_{\text{H}} = 2.28\times10^{20}$ cm$^{-2}$ \citep{2016A&A...594A.116H}. For models with redshift as a parameter, we froze it to the value of the cluster ($z=0.354$). 
\\ \indent When analyzing the thermodynamic properties of the ICM, the spectra extracted from the 15 OBSIDs were jointly fitted with a combination of different models:
\begin{itemize}[noitemsep,nolistsep,leftmargin=*]
\setlength{\parskip}{0pt}
\item The model \texttt{tbabs$\ast$apec} is composed of a photoelectric absorption model (\texttt{tbabs}), convolved with a thermal model (\texttt{apec}). The column density and redshift were fixed, while the other parameters (temperature $kT$, abundance and normalization) were left free to vary.
\item The model \texttt{projct$\ast$tbabs$\ast$apec} combines the previous model with the component \texttt{projct}, that computes the combined spectra of emission from a set of nested shells projected into annular regions. The inclusion of this component allows us to derive the deprojected electron density $n_{\text{e}}$ of the ICM by combining the normalization ($norm$) of the \texttt{apec} component with the volume (\textit{V}) of the emitting region. Assuming $n_{\text{e}}=1.2n_{\text{H}}$, $n = n_{\text{e}} + n_{\text{H}} = 1.83\,n_{\text{e}}$ (where $n$ and $n_{\text{H}}$ are the total density and the proton density, respectively), the electron density can be estimated as:
\begin{equation}
    \label{normne}
    n_{\text{e}} = \sqrt{10^{14}\left(\frac{4\pi \times norm \times[D_{\text{A}}(1+z)]^{2}}{V/1.2}\right)},
\end{equation} 
where $D_{\text{A}}$ is the angular diameter distance (1026 Mpc for RBS~797). The electron density can then be combined with the deprojected temperature of the \texttt{apec} component to obtain the pressure $p$, the entropy $S$ and the cooling time $t_{\text{cool}}$ of the ICM, defined as: 
\begin{align}
    \label{press}
        &p=nkT \\
    \label{entro}
        &S =\frac{kT}{n_{\text{e}}^{2/3}} \\
    \label{coolt}
        &t_{\text{cool}} = \frac{\gamma}{\gamma -1} \frac{kT}{\mu \,X \,n_{\text{e}}\,\Lambda(T)}
\end{align}
where $\gamma = 5/3$ is the adiabatic index, $\mu\approx0.6$ is the mean molecular weight, $X\approx0.7$ is the hydrogen mass fraction and $\Lambda(T)$ is the cooling function \citep{1993ApJS...88..253S}.
\end{itemize}

\subsection{Radio - VLA} \label{subsec:radio}
To allow a comparison with the X-ray data, we employ the archival radio observations of RBS~797 performed with the VLA at 1.4 GHz originally presented by \citet{2006A&A...448..853G} and \citet{2012ApJ...753...47D}. In particular, we show the radio contours generated from the array A observation (rms$=$0.02 mJy/beam, $\sim$1.5$''$ resolution; see figure 2b in \citealt{2006A&A...448..853G}), that best emphasize the structure of the radio galaxy and its interaction with the ICM in the inner $\sim$50 kpc of the cluster. Additionally, to investigate the radio/X-ray interaction at larger scales ($\sim$100 kpc), we show the radio contours obtained from the combined A-B-C array observations (rms$=$0.01 mJy/beam, $\sim$3$''$ resolution; see \citealt{2012ApJ...753...47D}).
We also show in \S\ref{subsec:epfeed} the preliminary contours of our new JVLA observations at 3 GHz, whose detailed analysis is still under way and will be presented in a forthcoming paper (Ubertosi et al., in prep).

\section{Shock fronts in the ICM} \label{sec:result}

The top panel of Fig. \ref{fig:fig1} shows the background subtracted, exposure corrected \textit{Chandra} image. In the bottom panels we show the original image filtered with the Gaussian Gradient Magnitude (GGM) filter \citep{2016MNRAS.460.1898S} with a 2$''$ filter, and the residual \textit{Chandra} image (originally presented in \citealt{2021ApJ...923L..25U}, obtained by subtracting from the \textit{Chandra} image a 2D double $\beta$-model). 
In the inner $\sim$50 kpc there are the four equidistant X-ray cavities already discussed in \citet{2021ApJ...923L..25U}, best visible in the residual image (right panel). Beyond the X-ray cavities, the images reveal the presence of three distinct and nested surface brightness edges, at approximate projected distances from the center of $\sim$50 kpc, $\sim$80 kpc and $\sim$130 kpc. In the following, we refer to the three features as \textit{inner}, \textit{middle} and \textit{outer} edges. 
\\ \indent The inner edge surrounds the AGN-inflated cavities, and consists of an ellipse with P.A.$\approx$345$^{\circ}$ and ellipticity (ratio between major and minor axis) of $\sim$1.2. The edge is located at a distance along the major axis of $\sim$11$''$ (54 kpc), and appears stronger along the E-W direction. Within this shell of bright X-ray emission there are the four cavities (at a distance of $\sim$5.5$''$ from the center, see \citealt{2021ApJ...923L..25U}) and the surrounding rims (at a distance of $\sim8''$ from the center). The middle edge appears sharper in the N-S direction, and is described by an ellipse with P.A. $\sim$20$^{\circ}$ and ellipticity 1.1. The outer edge is described by an ellipse with P.A. $\sim$100$^{\circ}$ and ellipticity 1.1, and is particularly pronounced in the E-W direction.
While the inner and middle edges were already noted by \citet{2011ApJ...732...71C}, the deeper \textit{Chandra} observations allowed us to recover the third outer edge. Most importantly, thanks to the higher number of counts we are able to perform a thorough morphological and spectral analysis of such edges for the first time (see \S\ref{subsec:wshocks}). 

\subsection{Detailed properties of the shocks}\label{subsec:wshocks}
Our aim is to secure the identification and further investigate the properties of the edges visible in Fig. \ref{fig:fig1}. In particular, this requires one [i] to identify the exact position and magnitude of each front by studying the surface brightness profile across the edge and [ii] to measure thermodynamic properties (e.g., temperature and pressure) inside and outside the front to determine its nature. In the following we first describe the procedure we employed to search for and investigate shock fronts, and then we present the results for RBS~797.
\\ \indent  We performed a systematic search for edges in the ICM by extracting surface brightness profiles (centered on the X-ray centroid, that coincides with the AGN in the BCG) in circular or elliptical sectors of varying opening angles (between 30$^{\circ}$ and 90$^{\circ}$) and different binning (between 0.7$''$ to 2$''$). This strategy was adopted to determine the geometry that best describes the fronts. The resulting profiles were visually inspected to identify possible jumps in surface brightness. For each edge, the profile was fit in \texttt{Proffit} with a single power-law model and with a broken power-law model: a surface brightness edge at distance $r_{J}$ and characterized by a density jump $J$ was considered a detection if an F-test between the single and the broken power-law indicated a significant statistical improvement (more than 99\% confidence). For the detected surface brightness edge, the Mach number $\mathcal{M}$ of the front was derived from the best-fit density jump $J$ using the Rankine - Hugoniot conditions (see e.g., \citealt{2007PhR...443....1M}):
\begin{equation}
    \label{machn}
    \mathcal{M} = \left( \frac{3J}{4-J} \right)^{1/2}
\end{equation}
Moreover, the Mach number can be used to predict the expected temperature and pressure jumps using again the Rankine - Hugoniot conditions:
\begin{align}
    \label{machkt}
    &T_{\text{jump}}^{\text{exp}} = \frac{5\mathcal{M}^{4} + 14\mathcal{M}^{2} - 3}{16\mathcal{M}^{2}} \\
    \label{machp}
    &p_{\text{jump}}^{\text{exp}} = \frac{5\mathcal{M}^{2} - 1}{4}
\end{align}
\\ \indent To measure the spectral properties of the detected surface brightness jumps, we extracted the spectra of three concentric regions: the first region is a wedge extending between 0.75$r_{J}$ - $r_{J}$; the second is a wedge extending between $r_{J}$ - 1.5$r_{J}$; the third wedge extends from 1.5$r_{J}$ to the edge of the \textit{Chandra} image, and allows for deprojection. These bin widths were chosen to avoid the inclusion of thermal emission far from the jump: while selecting larger regions would increase the number of counts, it may also lead to smearing thermodynamic gradients. Spectra were fitted with a \texttt{projct$\ast$tbabs$\ast$apec} model to measure the deprojected temperature and density (Eq. \ref{normne}), which were combined to derive the pressure jump across each edge (Eq. \ref{press}). 
\begin{figure*}[ht]
	\centering
	\gridline{\fig{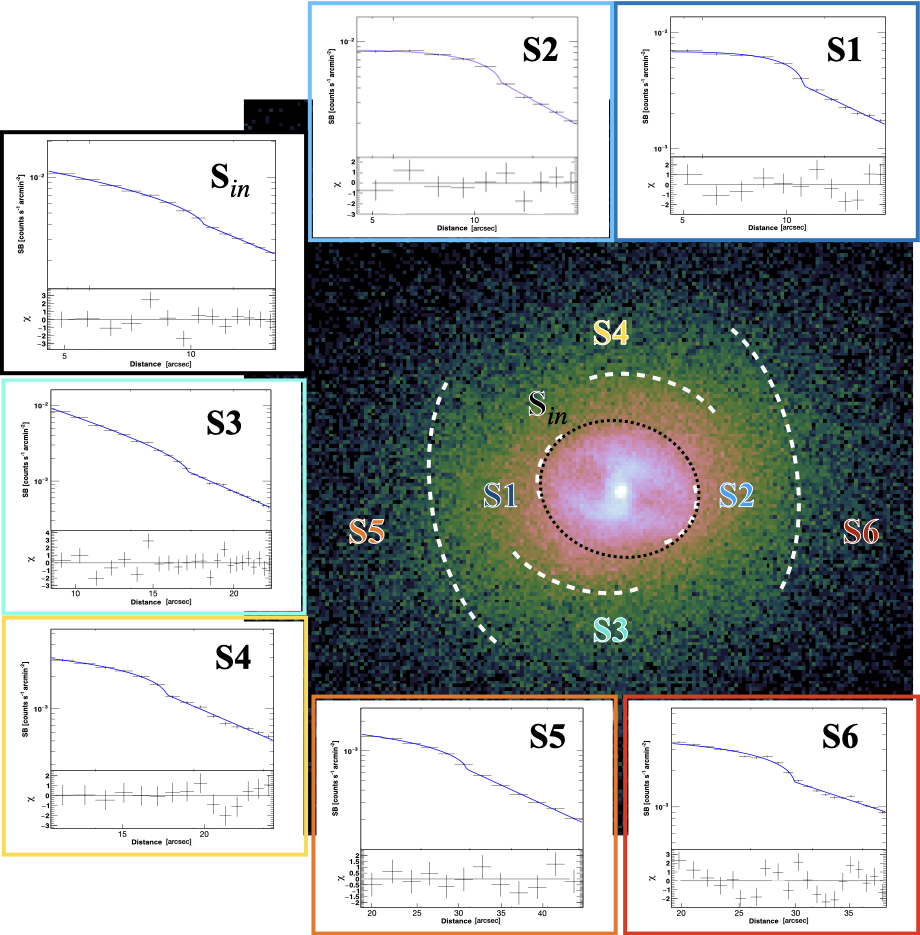}{1\linewidth}{}}
	\caption{\textit{Central panel:} 0.5 - 7 keV \textit{Chandra} image of RBS~797. The white (black) dashed regions indicate the positions and extents of the arc-like (cocoon-like) edges found on opposed sides of the center, as results from the surface brightness analysis reported in Tab. \ref{tab:shockbri}. Colored labels indicate the name of each edge. \textit{Sub-panels:} Surface brightness profile across each edge fit with the broken power-law model (blue line) described in Tab. \ref{tab:shockbri}. Residuals of the fit are shown in the bottom box of each plot.}
	\label{fig:surbr}
\end{figure*}
\\ \indent Applying the above procedure to RBS~797 allowed us to confirm the shock front nature of the three concentric and nested systems of edges noted in the previous paragraphs. In Fig. \ref{fig:surbr} we show the arc region that best describes each front and the fitted surface brightness profiles, while in Tab. \ref{tab:shockbri} of the Appendix and Tab. \ref{tab:specfronts} we report the surface brightness analysis and thermodynamic properties of the shocks, respectively. The details of the spectral fitting are shown in Tab. \ref{tab:fronts} and Fig. \ref{fig:shockreg} of the Appendix, while below we report the main properties of each detected shock: 
\linebreak
\\ \textit{Inner edges (S1, S2, S$_{\text{in}}$) -} The innermost surface brightness edge in RBS~797 is the bright cocoon surrounding the four X-ray cavities. By extracting the surface brightness profile in elliptical sectors (with ellipticity 1.2 and P.A. 345$^{\circ}$) we recovered sharp and significant density jumps between angles 130$^{\circ}$-190$^{\circ}$ and 310$^{\circ}$-15$^{\circ}$, i.e. beyond the E - W cavities (first two rows in Tab. \ref{tab:shockbri}, S1 and S2 panels in Fig. \ref{fig:surbr}). The edges are located at roughly 10$''$-11$''$ ($\approx$50 - 55 kpc) from the center. We name these inner east and inner west edges S1 and S2, respectively. By substituting in Eq. \ref{machn} the fitted density jumps we measured Mach numbers $\mathcal{M}_{\text{S1}} = 1.33\pm0.05$ and $\mathcal{M}_{\text{S2}} = 1.27\pm0.04$. The spectral analysis confirms the shock front nature of the edges (first two rows of Tab. \ref{tab:specfronts}): S1 is characterized by a temperature jump $T_{\text{jump}}^{\text{obs}}=1.35\pm0.18$ and a pressure jump $p_{\text{jump}}^{\text{obs}}=1.80\pm0.24$. The front S2 is less pronounced, being associated with a temperature jump of $T_{\text{jump}}^{\text{obs}}=1.13\pm0.08$ and $p_{\text{jump}}^{\text{obs}}=1.50\pm0.13$. While the lower Mach number of S2 compared to S1 is consistent with a less pronounced shock to the west, it is also possible that this is due to projection effects: low temperature gas in the W cavity rims, projected in front of the S2 front, may be damping the temperature gradient (similarly to what has been found for Hydra A, see \citealt{2011ApJ...732...13G}). \\ \indent  As the geometry of the edge is suggestive of a cocoon shock propagating in the ICM, we also analyzed the radial profile extracted from a complete ellipse with ellipticity 1.2 and P.A. $345^{\circ}$. We found a density jump (named S$_{\text{in}}$) at $\sim$10.7$''$ (52.4 kpc) from the center with a Mach number $\mathcal{M}_{\text{S}_{\text{in}}} = 1.20\pm0.03$. With the spectral analysis we found a temperature jump of $T_{\text{jump}}^{\text{obs}}=1.18\pm0.07$ and a pressure jump $p_{\text{jump}}^{\text{obs}}=1.88\pm0.16$ (see Fig. \ref{fig:surbr}, and third row of Tab. \ref{tab:shockbri} and Tab. \ref{tab:specfronts}). Thus, the inner edge is continuous around the azimuth, and consists of a cocoon shock driven by the jets of the central AGN. The lower Mach number of S$_{\text{in}}$ with respect to S1 and S2 probably indicates that the shock strength varies with azimuth (as also found e.g., in Hydra A, \citealt{2011ApJ...732...13G}), so $\mathcal{M}_{\text{S}_{\text{in}}}$ is the result of an average across the azimuth. Thus, we suggest that S1 and S2 are the highest Mach number parts of a single shock (S$_{\text{in}}$). In this respect, the ellipticity of S$_{\text{in}}$ (1.2) provides an indication of the ratio of the mean shock speeds along the major and minor axes.
\linebreak
\\ \textit{Middle edges (S3, S4) -} The second front identified from the images in Fig. \ref{fig:fig1} is located at a distance of 16$''$-17$''$ ($\approx$80 kpc) from the center. Extracting surface brightness profiles from elliptical sectors (ellipticity 1.1 and P.A. 20$^{\circ}$) revealed two significant jumps south of the center (S3, angles 200$^{\circ}$-270$^{\circ}$, Mach number $\mathcal{M}_{\text{S3}} = 1.24\pm0.03$) and north of the center (S4, angles 45$^{\circ}$-110$^{\circ}$, Mach number $\mathcal{M}_{\text{S4}} = 1.30\pm0.08$). The spectral analysis revealed that the shock front S3 has a temperature jump $T_{\text{jump}}^{\text{obs}}=1.23\pm0.11$ and a pressure jump $p_{\text{jump}}^{\text{obs}}=1.67\pm0.21$, while the shock front S4 has a temperature jump $T_{\text{jump}}^{\text{obs}}=1.24\pm0.12$ and a pressure jump $p_{\text{jump}}^{\text{obs}}=2.36\pm0.29$ (see Fig. \ref{fig:surbr}, Tab. \ref{tab:shockbri} and Tab. \ref{tab:specfronts}). We tested whether S3 and S4 may be part of a single front over 360$^{\circ}$ as done for the inner edge, but the surface brightness profile from the complete ellipse did not reveal any significant jump. 
\linebreak
\\ \textit{Outer edges (S5, S6) -} The presence of a third set of fronts at a distance of $\approx27''$ (or 135 kpc) from the center was not noticed in previous studies of RBS~797. Outer edges in surface brightness to the E-W are visible in the images of Fig. \ref{fig:fig1}. Fitting surface brightness profiles to the E-W in elliptical sectors (ellipticity 1.1 and P.A. 102$^{\circ}$) returned evidences for prominent density jumps at the location of the edges (between angles 140$^{\circ}$-252$^{\circ}$ and 304$^{\circ}$-87$^{\circ}$; see Fig. \ref{fig:surbr} and last two rows of Tab. \ref{tab:shockbri} and Tab. \ref{tab:specfronts}). In particular, we find that the eastern outer front (S5) has a Mach number $\mathcal{M}_{\text{S5}} = 1.19\pm0.03$ and coincides with jumps in temperature and pressure of $1.37\pm0.26$ and $1.91\pm0.41$, respectively. The western outer front is traveling with a Mach number $\mathcal{M}_{\text{S6}} = 1.25\pm0.04$ and coincides with jumps in temperature and pressure of $1.41\pm0.24$ and $2.34\pm0.43$, respectively. 
\linebreak
\\ \indent Overall, the above results confirm the existence of three groups of nested shock fronts in the ICM of RBS~797. The significance of the measured density jumps $J$ is more than $\sim$6$\sigma$ (see Tab. \ref{tab:shockbri}). For the inner edge and the middle edge, the significance of the measured temperature jumps is between 2-3$\sigma$, while the significance of the pressure jumps is between 3-5$\sigma$. For the outer edge, the significance of temperature and pressure jumps is between 1.5-3$\sigma$. Moreover, the predicted and measured temperature and pressure jumps are in agreement within errors (see Tab. \ref{tab:specfronts}, fourth to seventh columns). With Mach numbers in the range 1.2 - 1.3 the edges can be classified as weak shocks, likely resulting from successive energy injection events by the AGN activity. It is interesting to note the continual change in position angle of the different shock groups: from E-W (inner shocks) to N-S (middle shocks) and again to E-W (outer shocks). It is likely that the middle and outer opposed shock arcs represent the highest Mach number parts of complete middle and outer cocoon shocks that encompass the whole azimuth, respectively, as we were able to verify for the inner jump. For instance, we might not be able to detect \textit{middle E-W} surface brightness jumps as these would be located in a rather narrow region in between the inner and outer E-W fronts, which would prevent the surface brightness profile at the interface of the shocks from being accurately measured. Besides, it is known that the central AGN in RBS~797 is characterized by multiple changes in position angle of its lobes and jets \citep{2006A&A...448..853G,2013A&A...557L..14G,2011ApJ...732...71C,2021ApJ...923L..25U}, up to 90$^{\circ}$ misalignment. Thus, the nested groups of opposed and misaligned shock fronts might trace older episodes of differently oriented AGN activity cycles. We further investigate this scenario in \S\ref{sec:disc}.
\setlength{\tabcolsep}{3.5pt}
\begin{table*}
	\centering
	\caption{Properties of the nested shocks in RBS~797.}
	\label{tab:specfronts}
		\begin{tabular}{l|cccc|cc|ccc}
			\hline
			Shock & $r_{\text{sh}}$ &$\mathcal{M}$ & $T_{\text{jump}}^{\text{exp}}$ & $p_{\text{jump}}^{\text{exp}}$& $T_{\text{jump}}^{\text{obs}}$ & $p_{\text{jump}}^{\text{obs}}$ &  t$_{\text{\text{age}}}$& E$_{\text{sh}}$ & P$_{\text{sh}}$ \\
		     & [kpc ($''$)] &  & & &  &  &  [Myr] & [10$^{60}$ erg] & [10$^{45}$ erg s$^{-1}$] \\
			\hline
			
			\rule{0pt}{4ex} S1 - Inner East & 53.9 (11.0) & 1.33$\pm$0.05 &  1.33$\pm$0.05 & 1.98$\pm$0.17 & 1.35$\pm$0.18 & 1.80$\pm$0.24 & 28.5$\pm$2.4 & 8.3$\pm$1.7& 8.6$\pm$1.8\\

			\rule{0pt}{4ex} S2 - Inner West & 50.5 (10.3) & 1.27$\pm$0.04 & 1.26$\pm$0.04 & 1.77$\pm$0.08 & 1.13$\pm$0.08 & 1.50$\pm$0.13 & 32.8$\pm$2.8 & 4.2$\pm$0.8 & 4.1$\pm$0.8\\
			
			\rule{0pt}{4ex} S$_{\text{in}}$ - Inner Total & 52.4 (10.7) & 1.20$\pm$0.03 & 1.19$\pm$0.02 & 1.54$\pm$0.06 & 1.18$\pm$0.07 & 1.88$\pm$0.16 & 33.4$\pm$1.3 & 6.6$\pm$0.5& 6.3$\pm$0.6\\
			
			\hline
			
			\rule{0pt}{4ex} S3 - Middle South & 79.4 (16.2) & 1.20$\pm$0.03 & 1.19$\pm$0.03 & 1.54$\pm$0.08 & 1.23$\pm$0.11 & 1.67$\pm$0.21 & 50.9$\pm$4.3 & 8.0$\pm$1.9& 5.0$\pm$1.2\\
			
			\rule{0pt}{4ex} S4 - Middle North & 81.8 (16.7) & 1.19$\pm$0.03 & 1.18$\pm$0.03 & 1.51$\pm$0.10 & 1.24$\pm$0.12 & 2.36$\pm$0.29 & 54.9$\pm$4.5 & 6.3$\pm$1.5& 3.7$\pm$0.8\\
			
			\hline
			
			\rule{0pt}{4ex} S5 - Outer East & 136 (27.7) & 1.19$\pm$0.03 & 1.18$\pm$0.03 & 1.49$\pm$0.08 & 1.37$\pm$0.26 & 1.91$\pm$0.41 & 84.7$\pm$6.4 & 18.5$\pm$5.5& 7.0$\pm$2.1\\
			
			\rule{0pt}{4ex} S6 - Outer West & 129 (26.3) & 1.25$\pm$0.04 & 1.24$\pm$0.03 & 1.70$\pm$0.10 & 1.41$\pm$0.24 & 2.34$\pm$0.43 & 82.8$\pm$6.9 & 19.9$\pm$3.0& 7.7$\pm$1.5\\
		
			\hline
		\end{tabular}
		\tablecomments{(1) Shock label; (2) distance of the shock from the center (measured from the front mid-aperture); (3) Mach number of the shock, obtained from the density jump reported in Tab. \ref{tab:shockbri}; (4 - 5) temperature and pressure jumps predicted by the Mach number; (6 - 7) observed temperature (deprojected) and pressure jumps, obtained from the fit to the spectrum of the ICM in the downstream and upstream sides of the edge (see Tab. \ref{tab:fronts}); (8) age of the shock (see Eq. \ref{shockage}); (9) energy of the shock (see Eq. \ref{shockenergymach}); (10) shock power, defined as $P_{\text{sh}} = E_{\text{sh}}/t_{\text{age}}$. The energy and power of fronts S1, S2, S3, S4, S5 and S6 are referred to the half ellipsoid covered by each front, while for S$_{\text{in}}$ the full ellipsoid was considered (see text for details).}
\end{table*}
\subsection{Shock energetics and timescales}
Measuring the age and energetics of the shocks provides essential information for probing the impact of these features on the cluster thermodynamic conditions. The age of the shocks $t_{\text{age}}$ can been determined by assuming that the shock has traveled from the center to its current position ($r_{\text{sh}}$, the distance to the mid-aperture of the front, reported in Tab. \ref{tab:specfronts}) with its observed Mach number $\mathcal{M}$, i.e.:
\begin{equation}
    \label{shockage}
    t_{\text{age}} = \frac{r_{\text{sh}}}{\mathcal{M}\,c_{\text{S}}}
\end{equation}
where $c_{\text{S}}=\sqrt{\gamma kT/(\mu m_{\text{p}})}\approx5.2\times10^{2}\sqrt{kT[\text{keV}]}$ km s$^{-1}$ is the upstream sound speed (measured from the temperature outside the shock). We note that this method may slightly overestimate the true shock age (by a relatively modest factor of $\sim10-20\%$, see \citealt{2011ApJ...726...86R}), given that when the shock was initially launched it likely had a higher Mach number. Using Eq. \ref{shockage} we found an age of 33.4$\pm$1.3 Myr for the inner shock $S_{\text{in}}$, 52.9$\pm$2.8 Myr for the middle shock (estimated as the average between $S3$ and $S4$), and 83.8$\pm$1.4 Myr for the outer shock (average between $S5$ and $S6$). See Tab. \ref{tab:specfronts}) for details. 
\\ \indent To compute the energy deposited by each shock we considered the volume of shocked gas and the difference in energy density at the interface (e.g. \citealt{2001ApJ...557..546D,2015ApJ...805..112R}):
\begin{equation}
    \label{shockenergy}
    E_{\text{sh}} = \frac{3}{2}\,V \times \Delta p = \frac{3}{2}\,V \times (p_{\text{in}} - p_{\text{out}})
\end{equation}
The above equation can be rewritten in terms of pre-shock (upstream) pressure $p_{\text{out}}$, shock volume $V$ and Mach number $\mathcal{M}$ as:
\begin{equation}
    \label{shockenergymach}
    E_{\text{sh}} = \frac{3}{2} \,p_{\text{out}} \times V \times \left(\frac{5\mathcal{M}^{2}-5}{4}\right)
\end{equation}
The pressure inside (downstream) and outside (upstream) each front is known from the spectral analysis of the shocks (see Tab. \ref{tab:fronts} in the Appendix). For each front of each shock pair (S1 and S2, S3 and S4, S5 and S6) we computed the volume as that of one half of a prolate ellipsoid (assuming that the two fronts of each pair are parts of a single cocoon) of major and minor axes set by the annular sector used for the surface brightness and spectral analysis (see Tab \ref{tab:fronts}; for comparison, see also \citealt{2011ApJ...726...86R,2018ApJ...855...71S}). For the cocoon front S$_{\text{in}}$, we considered the full prolate ellipsoid. The shock energies are reported in Tab. \ref{tab:specfronts}. For the inner edge we have two estimates of the total shock energy, i.e. $E_{\text{S1}} + E_{\text{S2}} = 12.5\pm2.5\times10^{60}$ erg or $E_{\text{S}_{\text{in}}} = 6.6\pm0.3\times10^{60}$. The first method likely represents an upper limit to the true shock energy, as it assumes that the whole shock front has the Mach number of its strongest parts. Thus, we consider $E_{\text{S}_{\text{in}}}$ as our best estimate for the shock energy. For the middle and outer shocks we compute the total energy as $E^{\text{m}}_{\text{sh}} = E_{\text{S3}} + E_{\text{S4}}$ and $E^{\text{o}}_{\text{sh}} = E_{\text{S5}} + E_{\text{S6}}$, respectively, which may overestimate the true shock energy by a factor up to 2 (based on the comparison with the inner edge).  
\\ \indent We thus find that the energy rises from the inner shock ($E^{\text{in}}_{\text{sh}} \sim 0.7\times10^{61}$ erg), to the middle shock ($E^{\text{m}}_{\text{sh}} \sim 1.4\times10^{61}$ erg), to the outer shock ($E^{\text{o}}_{\text{sh}} \sim 3.8\times10^{61}$ erg), mostly due to the larger volume occupied by the progressively more distant fronts (see also \citealt{2015ApJ...805..112R}). These values can be summed up to obtain the total energy injected through shocks in the ICM of RBS~797, that is E$^{tot}_{\text{sh}} = 5.9\pm1.5\times10^{61}$ erg.  
\\ \indent Ultimately, we computed the shock power as the ratio between the shock energy and the shock age, $P_{\text{sh}}=E_{\text{sh}}/t_{\text{sh}}$. This information is crucial to understand how effective the energy injection by shocks is with respect to other form of energy inputs (e.g. X-ray cavities) or losses (i.e. radiative cooling), a topic we discuss in \S\ref{sec:disc}. On the one hand, the shock energy computed using Eq. \ref{shockenergymach} may be overestimated by up to a factor of $\sim$2 (see \citealt{2011ApJ...726...86R}). On the other hand, as stated above the shock ages may also be slightly overestimated. Thus, in computing the shock power as $E_{\text{sh}}/t_{\text{sh}}$, these two effects partially compensate for one another. We report the shock powers in Tab. \ref{tab:specfronts}; we estimate total powers of $P^{\text{in}}_{\text{sh}} = 6.3\pm0.6\times10^{45}$ erg s$^{-1}$, $P^{\text{m}}_{\text{sh}} = 8.7\pm1.9\times10^{45}$ erg s$^{-1}$ and $P^{\text{o}}_{\text{sh}} = 1.5\pm0.4\times10^{46}$ erg s$^{-1}$. Since the middle and outer shock powers are based on the highest Mach number parts of the shocks, the true $P^{\text{m}}_{\text{sh}}$ and $P^{\text{o}}_{\text{sh}}$ may be slighly lower. It is then possible to conclude that, within errors, the shock power has remained nearly constant, releasing for every outburst roughly $(0.6 - 1.5)\times10^{46}$ erg s$^{-1}$ in the ICM.

\section{Global properties of the ICM}\label{ssec:genicm}
The following subsections are dedicated to the analysis of the global properties of the ICM in RBS~797. After deriving the radial profiles (\S\ref{subsec:radia}) and maps (\S\ref{subsec:spect}) of thermodynamic quantities, we study the abundance distribution in the ICM (\S\ref{subsect:metal}).

\subsection{Radial profiles of thermodynamic properties} \label{subsec:radia}
\begin{figure*}[ht]
	\centering
	\gridline{\fig{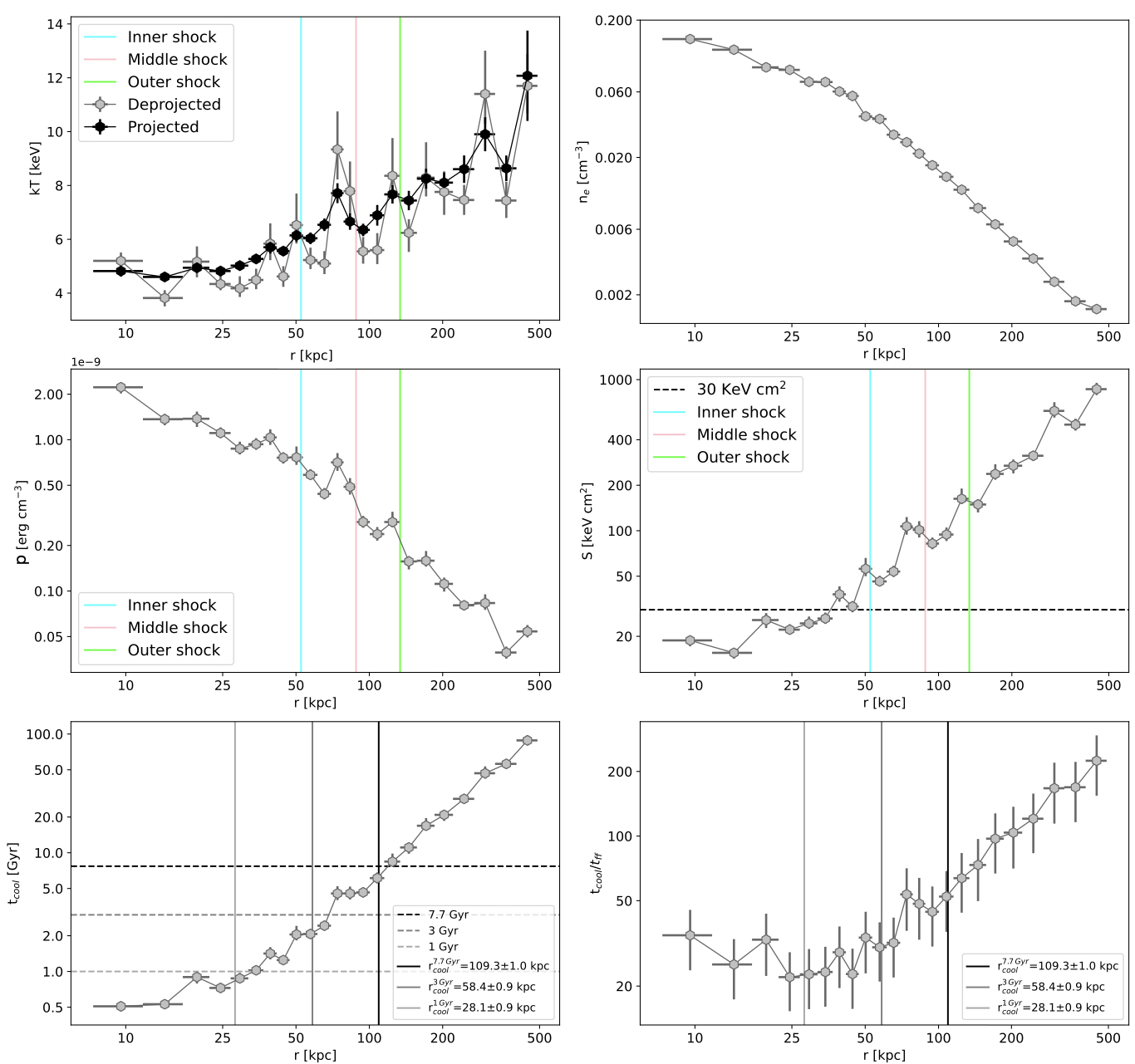}{1\linewidth}{}}
	\caption{Radial profiles of ICM thermodynamic quantities in RBS~797. \textit{Upper left:} projected (black dots) and deprojected (grey dots) temperature profiles; \textit{Upper right:} electron density profile; \textit{Middle left:} pressure profile; \textit{Middle right:} entropy profile, with the horizontal dotted line showing the $S\leq30$ keV cm$^{2}$ threshold for the condensation of the ICM into multi-phase gas clouds; \textit{Lower left:} cooling time profile, with horizontal lines showing different indicators (1 Gyr, 3 Gyr and 7.7 Gyr) of the cool core region, the extent of which for each indicator is marked by the vertical lines; \textit{Lower right:} radial profile of the ratio between the cooling time and the free-fall time ($t_{\text{cool}}/t_{\text{ff}}$), with the different estimates for the cooling radius overlaid with vertical lines. In the temperature, pressure and entropy profiles the colored vertical lines show the distance from the center of the three nested weak shocks (see \S\ref{sec:result}). For details on how these profiles were built see \S\ref{subsec:radia}.}
	\label{fig:therm}
\end{figure*}
Radial profiles of thermodynamic variables (temperature, density, pressure, entropy and cooling time) are essential tools for investigating cooling, feedback, and the thermodynamic state of clusters. In order to build such profiles, we extracted the spectra of circular annuli centered on the AGN and extending between 1.5$''$ - 100$''$ (7.5 kpc - 500 kpc). The width of each annulus is designed to obtain more than 8000 counts in the 0.5 - 7 keV band, enabling us to constrain temperatures with an accuracy of 5\% (8\% when deprojected). 
The spectra were fit with a projected (\texttt{tbabs$\ast$apec}) and a deprojected (\texttt{projct$\ast$tbabs$\ast$apec}) thermal model.
\\ \indent We show the resulting radial profiles of temperature (projected in black and deprojected in gray), density, pressure, entropy and cooling time (estimated using Eq. \ref{normne},\ref{press},\ref{entro},\ref{coolt}) in the first five panels of Fig. \ref{fig:therm}. We verified that neither intrinsic absorption (\texttt{ztbabs$\ast$tbabs$\ast$apec}) or an additional thermal component (\texttt{tbabs$\ast$(apec+apec)}) is required in any of the annuli. We also checked that using the \texttt{DSDEPROJ} code \citep{2007MNRAS.381.1381S,2008MNRAS.390.1207R} to derive deprojected quantities returns consistent results with those obtained by using \texttt{projct} (see Fig. \ref{fig:app-pro}).
\begin{figure*}[ht]
	\centering
	\gridline{\fig{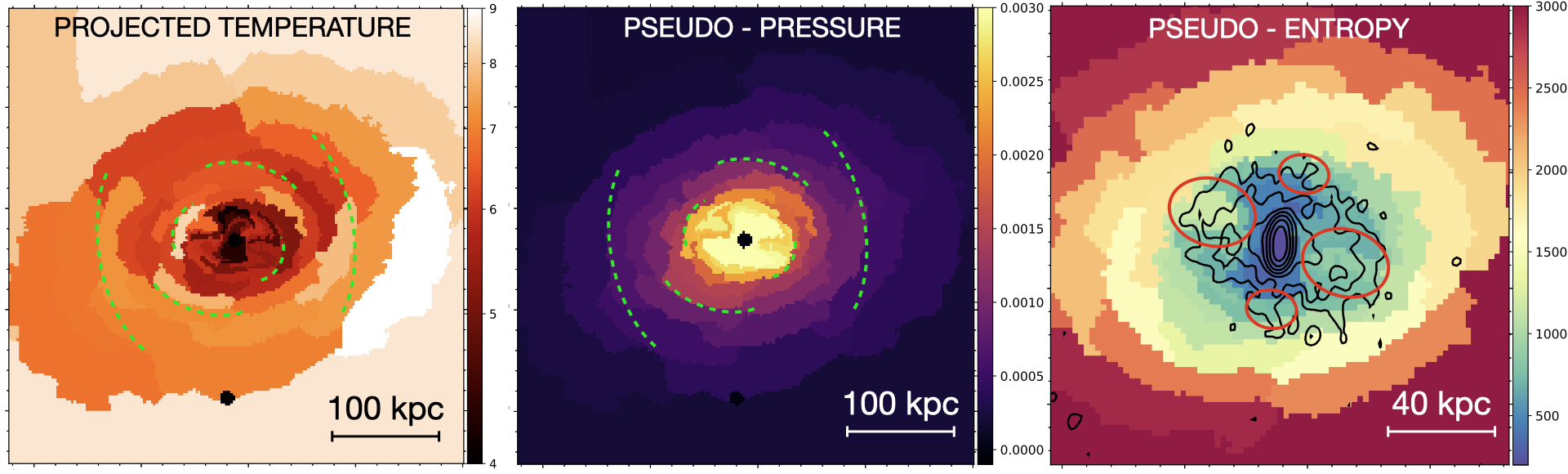}{1\linewidth}{}}
	\vspace{-0.5cm}
	\caption{Maps of projected temperature (in keV, \textit{left panel}), pseudo-pressure (in arbitrary units, \textit{middle panel}) and pseudo-entropy (in arbitrary units, \textit{right panel}). The spectrum extracted from each region has a SNR$\geq$60. Relative uncertainties on the mapped values are $\leq$20\%. In the \textit{left} and \textit{middle} panels, the green dashed arcs indicate the position and extent of the arc-like edges identified in \S\ref{sec:result}. In the \textit{right panel} the black contours show the radio galaxy at 1.4 GHz (see \citealt{2006A&A...448..853G}), while the red ellipses indicate the X-ray cavities. The central X-ray point source has been excluded from the spectral fitting.}
	\label{fig:spectralrbs}
\end{figure*}
\\ \indent The inwardly increasing density and decreasing temperature, entropy and cooling time confirm that RBS~797 is a cool core cluster (as already noticed by \citealt{2001A&A...376L..27S,2004ApJ...607..800B,2006A&A...448..853G,2011ApJ...732...71C,2012ApJ...753...47D}). With our deeper exposure, we are now able to resolve localized gradients in thermodynamic properties. We overplot on the temperature, pressure and entropy profiles the distance from the center of the three nested shocks. Both the projected and deprojected temperature profiles reveal an increase in temperature behind the three edges; moreover, the pressure and entropy within the edges are higher than outside.
\\ \indent The radial spectral analysis allows us to characterize the efficiency of ICM cooling in RBS~797. First of all, we aim at determining the extent of the cool core region, usually defined in the literature as the region where the gas has a cooling time less than 7.7 Gyr (corresponding to the look-back time of $z\sim1$ for the assumed cosmology; \citealt{2004ApJ...607..800B}). Other studies instead consider more conservative thresholds, as 3 Gyr or 1 Gyr (typical timescale from the last major merger; e.g. \citealt{2018ApJ...858...45M}). Thus, we fitted the cooling time profile with a power-law and located the intersection with t$_{\text{cool}} = 1,\,3,\,7.7$ Gyr. We find that the cooling radius in RBS~797 (where cooling time is less than 7.7 Gyr\footnote{We are aware that for RBS~797 ($z=0.354$) the time that has passed since $z=1$ is not 7.7 Gyr but $\sim$4 Gyr. However, a $t_{\text{cool}}\leq7.7$ Gyr is the typical approach of several studies in the literature (e.g., \citealt{2004ApJ...607..800B,2006ApJ...652..216R}), and has also been adopted for studies of clusters at higher redshift than RBS~797, up to $z\sim1.2$ (e.g., \citealt{2012MNRAS.421.1360H,2015ApJ...805...35H,2017MNRAS.471.1766B}). Thus, using this threshold - while possibly non-physical - enables us to draw comparison with the literature.} is r$_{\text{cool}}^{7.7\,Gyr}=109.$3$\pm$1.0 kpc (or 22.4$''$ - consistent with \citealt{2012ApJ...753...47D}). The cooling time falls below 3 Gyr within 58.4$\pm$0.9 kpc and below 1 Gyr within 28.1$\pm$0.9 kpc. Second of all, we wish to locate the radial range over which the ICM is not only cooling, but multi-phase gas is also expected to be present. To fulfill this aim, we employed different ICM cooling diagnostics. On the one hand, according to \citet{2016ApJ...830...79M} the gas should become multi-phase when the entropy is lower than 30 keV cm$^{2}$ or the cooling time falls below 1 Gyr. As it can be seen in Fig. \ref{fig:therm}, these conditions are both satisfied within roughly 30 kpc from the center. On the other hand, it has been suggested that cooling of the ICM into warm clouds occurs when the ratio between the cooling time and the free fall time (t$_{\text{ff}}$) is of the order of a few tens (10 - 30, e.g., \citealt{2015ApJ...799L...1V}). To measure the free fall time we derived the hydrostatic mass (e.g., \citealt{2006MNRAS.368..518V}) of RBS~797 from the deprojected pressure and density profile (as anticipated in \citealt{2021ApJ...923L..25U}). Then, from the mass profile $M(r)$ it is possible to obtain the free fall time profile as:
\begin{equation}
\label{tfreefall}
    t_{\text{ff}}(r) = \sqrt{\frac{2r^{3}}{GM(r)}}
\end{equation}
We show in Fig. \ref{fig:therm} the radial profile of $t_{\text{cool}}/t_{\text{ff}}$, which reveals that the ratio approaches $\sim$20-30 below $\sim$35 kpc from the center, which is also the extent of gas with $S\leq30$ keV cm$^{2}$ and $t_{\text{cool}}\leq1$ Gyr. Therefore, in RBS~797 we find that the conditions for the presence of multi-phase gas are met within a few tens of kpc from the center. \\ \indent We note that the azimuthally averaged radial analysis can predict the extent of the region where condensation should occur, but does not provide information on the azimuthal geometry of cold and dense gas. Thus, in the following subsection we show different methods for building maps of ICM thermodynamic quantities, that can provide valuable information on the radial and azimuthal variations of thermodynamic quantities and cooling efficiency.

\subsection{Thermodynamic maps of the ICM} \label{subsec:spect}
\begin{figure*}[ht]
	\centering
	\gridline{\fig{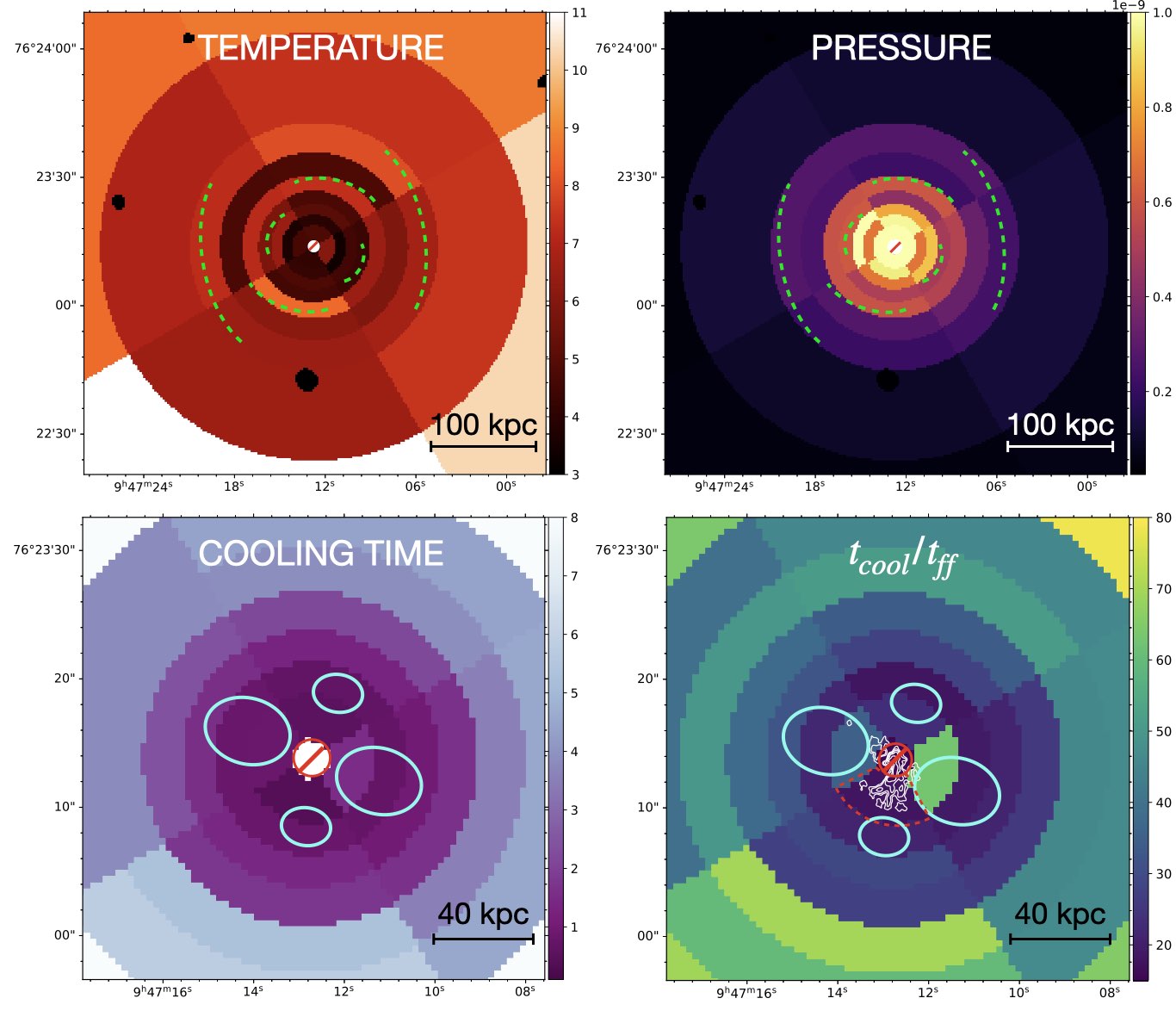}{1\linewidth}{}}
	\caption{
	Maps of deprojected temperature (\textit{upper left}, in keV), pressure (\textit{upper right}, in erg cm$^{-3}$), cooling time (\textit{lower left}, in Gyr) and $t_{\text{cool}}/t_{\text{ff}}$ ratio (\textit{lower right}) of the ICM, built by dividing the azimuth in four sectors and nine annuli (see text for details). In the \textit{upper} panels, the green dashed arcs indicate the position and extent of the arc-like edges identified in \S\ref{sec:result}. In the \textit{lower} panels, cyan ellipses show the position of the X-ray cavities. 
	In the \textit{lower right} panel, white contours from \citet{2022ApJ...940..140C} trace the morphology of the central [OII] emitting nebulae, and the red dashed wedge shows the region where the minimum in $t_{\text{cool}}/t_{\text{ff}}$ is found.
	Relative uncertainties in temperature, pressure and cooling time are on the order of $\sim$10\%, while those in $t_{\text{cool}}/t_{\text{ff}}$ are on the order of $\sim$20\%. The central X-ray point source has been excluded from the spectral fitting.}
	\label{fig:thermomap}
\end{figure*}
To enable high-resolution 2D mapping of ICM thermodynamic properties we built maps by using the contour binning technique (\texttt{CONTBIN}, \citealt{2006MNRAS.371..829S}) and setting a minimum signal-to-noise ratio (SNR) of 60. The spectrum extracted from each region was fit with a \texttt{tbabs$\ast$apec} model, leaving the temperature, metallicity and normalization free to vary. As done for the radial profile described in \S\ref{subsec:radia}, we verified whether the inclusion of intrinsic absorption or the fit with a two-temperature model could improve the description of the extracted spectra, but we found that no additional component is required. \\ \indent Then, by combining the temperature and normalization we derived maps of pseudo-pressure and pseudo-entropy in arbitrary units, using the following method: 
the \texttt{apec} normalization is proportional to the volume integral of the square of the electron density, i.e. $norm~\propto~\int ~n_{\text{e}}^{2}~dV$. Thus, the average projected electron density can be derived from the square root of the \texttt{apec} normalization scaled by a factor that accounts for the volume. We define this factor as $R\times n_{\text{pix}}$, where $R$ is the average distance of the spectral region from the center and $n_{\text{pix}}$ is the number of pixels in each spectral region (see also e.g., \citealt{2007A&A...463..839R,2011ApJ...737...99B,2015ApJ...805..112R}). Therefore, we measure the pseudo-electron density using the following proportion:
\begin{equation}
    \label{pseudo-density}
    n_{\text{e}} \propto \left( \frac{norm}{R\times n_{\text{pix}}} \right)^{1/2}
\end{equation}
From the pseudo-density it is trivial to measure pseudo-pressure and pseudo-entropy using Eq. \ref{press} and Eq. \ref{entro}. The resulting maps (shown in Fig. \ref{fig:spectralrbs}) show that the cooler and lower entropy gas is preferentially found in the bright central bar oriented north-south, and in the rims of the four X-ray cavities. These low entropy structures are likely the results of the combined expansion of the four X-ray cavities in the ICM, that may have uplifted the surrounding low entropy medium and/or triggered its condensation. By overlaying on the thermodynamic maps the arcs representing the shock fronts discussed in \S\ref{sec:result}, it is possible to appreciate the presence of high temperature and high pressure regions behind the arcs. Additionally, the entropy map shows that the low entropy gas encasing the X-ray cavities is more extended to the west. This supports the idea that low temperature gas projected in front of S2 may be damping the temperature gradient, as hypothesized in \S\ref{subsec:wshocks}.
\\ \indent Overall, the radial and thermodynamic maps analysis revealed that the ICM of RBS797 presents several radial and azimuthal gradients in spectral properties. In order to further investigate these features with improved statistics, and with the aim of measuring \textit{deprojected} thermodynamic quantities at different radii and orientations, we performed a spectral mapping in annular sectors. In particular, we divided the azimuth in four sectors with angular aperture of 90$^{\circ}$, starting in the west, with a center line at P.A. $345^\circ$ for the first sector (to
match the approximate symmetry of the central region). Then, each sector was divided in 9 annular regions, so that each annular sector contained at least the same number of counts (8000 in the 0.5 - 7 keV band) as the annuli used to construct the radial profiles in Fig. \ref{fig:therm}. By fitting the spectra with a \texttt{projct$\ast$tbabs$\ast$apec} model we are thus able to probe \textit{radial and azimuthal} variations of deprojected thermodynamic properties (see also \citealt{2010ApJ...714..758G}). To derive useful information on the distribution of cooling gas we also mapped the cooling time and the $t_{\text{cool}}/t_{\text{ff}}$ ratio, by measuring the free fall time at the radial distance of each annulus from the center (Eq. \ref{tfreefall}).\\ \indent 
The resulting deprojected temperature and pressure maps (see Fig. \ref{fig:thermomap}) highlight the concentric, nested and misaligned weak shocks (as found in the projected thermodynamic maps of Fig. \ref{fig:spectralrbs}). We note that due to the choice of circular symmetry, and the fact that the wedges do not exactly follow the shock fronts, the outer jump is less well defined. 
\\The cooling time and $t_{\text{cool}}/t_{\text{ff}}$ maps (lower panels in Fig. \ref{fig:thermomap}) confirm that short cooling time ($\leq$1 Gyr) gas is filling the space between and around the perpendicular radio lobes and cavities, supporting a strong connection between the AGN outbursts and ICM cooling. We note that the innermost southern wedge (red dashed region in Fig. \ref{fig:thermomap}) represents the locus where cooling should be most effective: we measure a local cooling time of $t_{\text{cool}}=438\pm51$ Myr and $t_{\text{cool}}/t_{\text{ff}} = 17.9\pm2.0$ (and an entropy $S=15.4\pm1.2$ keV cm$^{2}$, not shown here). We thus expect that filamentary multi-phase warm gas arising from condensation of the ICM should preferentially be found in this region. In this context, \citet{2011ApJ...732...71C} noted filamentary structures in the residual optical image of the BCG extending 8 - 10 kpc southward. These structures were recently confirmed by \citet{2022ApJ...940..140C}, who produced continuum-subtracted [O~II] maps of nebular emission using the \textit{Hubble Space Telescope}. The overlay of [O~II] contours on the $t_{\text{cool}}/t_{\text{ff}}$ map (see lower right panel in Fig. \ref{fig:thermomap}) reveals that cool gas at $\sim10^{4}$ keV is coincident with the region where the minimum in $t_{\text{cool}}/t_{\text{ff}}$ is found (within $\sim$20 kpc south from the center). These results may indicate that the local ICM is actually condensing into cool gas.

\subsection{Metallicity of the ICM}\label{subsect:metal}
\begin{figure}[ht]
	\centering
	\gridline{\fig{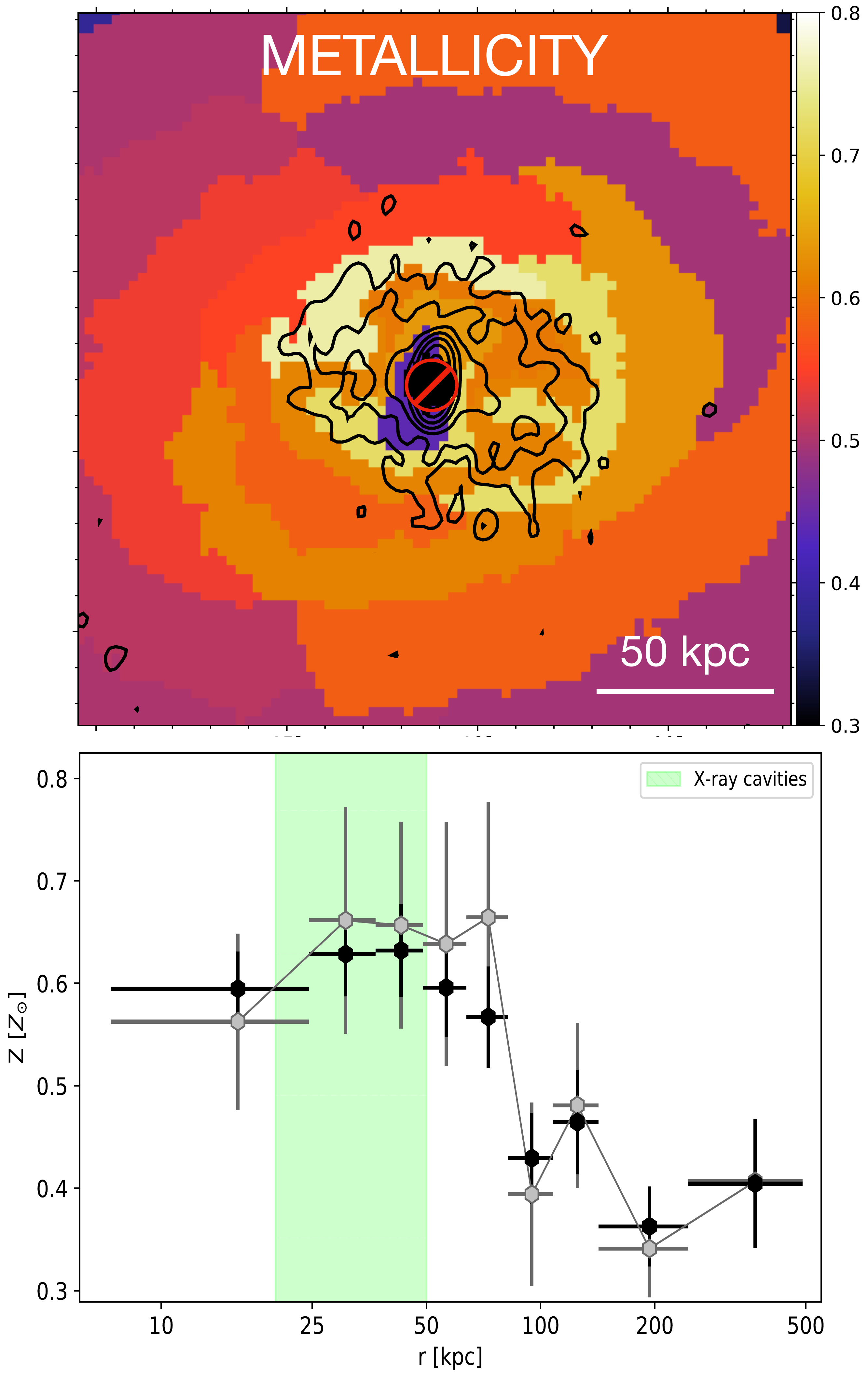}{1\linewidth}{}}
	\caption{\textit{Upper panel:} metallicity map of the ICM in $Z_{\odot}$. The spectrum extracted from each region has a SNR$\sim$100. Relative uncertainties on the mapped values are $\leq$10-15\%. Overlaid in black are the contours of the radio galaxy at 1.4 GHz (same as in Fig. \ref{fig:fig1}; \citealt{2006A&A...448..853G}). The central X-ray point source has been excluded from the spectral fitting. \textit{Lower panel}: projected (black dots) and deprojected (grey dots) metallicity radial profile, with the green region showing the radial range where the X-ray cavities are located.}
	\label{fig:metals}
\end{figure}

A previous study of the metallicity distribution based on the analysis of the $\sim$50 ks exposure found tentative indications of higher abundances in the direction of the E-W cavities compared to the surrounding medium \citep{2012ApJ...753...47D}, possibly suggesting that the thrust associated with the cavities' expansion was uplifting enriched gas from the center, as seen in a number of other systems (e.g. \citealt{2011ApJ...731L..23K,2015MNRAS.452.4361K}). Our aim is to follow-up on this argument with the deeper \textit{Chandra} exposure, by analyzing the radial and azimuthal distribution of metals in RBS~797. 
\\ \indent Constraining the abundances of the ICM with \textit{Chandra} typically requires spectra with a high SNR ($\gtrapprox100$). Therefore, to map the ICM abundances we built another set of maps requiring for each spectral extraction region a minimum SNR of 100 in the 0.5 - 7 keV band. Fitting the spectra with a \texttt{tbabs$\ast$apec} model allowed us to map abundances with relative uncertainties of $\leq$10-15\%. The metallicity map obtained with this choice of SNR is shown in Fig. \ref{fig:metals}. We note that within the inner $\sim$50-70 kpc the abundance is higher than outside $\sim$80 kpc from the center, which is expected in a cool core cluster. A peculiar feature in the metallicity map is the presence of a ring of enhanced metallicity at $\sim$50 kpc from the center. By overlaying 1.4 GHz radio contours on the map it is possible to see that this ring surrounds the lobes of the central radio galaxy to the east, north, west and south-west of the center. On the one hand, this feature is consistent with the scenario proposed by \citet{2012ApJ...753...47D} that the AGN is responsible for mechanical uplift of enriched gas from the center. On the other hand, the deep \textit{Chandra} exposure seems to rule out a bipolar uplift, as the ring covers almost the whole azimuth. In fact, the discovery of four equidistant cavities excavated by the AGN in perpendicular directions \citep{2021ApJ...923L..25U} is consistent with an azimuthally-symmetric enhancement. This has been also predicted by high-resolution hydrodynamical simulations of AGN jet feedback (e.g., \citealt{2011MNRAS.411..349G}).
\\ \indent  We note that both the analysis of \citet{2012ApJ...753...47D} and the metallicity map shown in Fig. \ref{fig:metals} do not take into account projection effects. Measuring accurate deprojected abundances requires an even larger SNR. Thus, to build a high-fidelity radial profile of abundances we extracted the spectrum from nine concentric circular annuli (extending between 1.5$''$ - 100$''$ from the center) centered on the AGN, requiring $\sim$30000 counts per bin (or SNR$\sim$200). By fitting again the spectra with a \texttt{tbabs$\ast$apec} model we verified that relative uncertainties on projected and deprojected abundances are of $\sim6\%$ and $\sim10-15\%$, respectively. The projected and deprojected profiles shown in Fig. \ref{fig:metals} confirm the general decreasing trends of abundances with radius; there is a clear gradient at roughly 80 kpc from the center, with average (deprojected) abundances of $\langle Z(r\leq80\,\text{kpc}) \rangle = 0.62\pm0.04 Z_{\odot} $ and of $\langle Z(r\geq80\,\text{kpc}) \rangle = 0.41\pm0.05 Z_{\odot}$. Furthermore, the projected and deprojected profiles hint at a slightly lower central abundance (by approximately 15$\%$) than in the region where the X-ray cavities are located (20 - 50 kpc, overplotted in green).
As a sanity check, we tested whether our results could be affected by the iron bias (see e.g., \citealt{1999MNRAS.309..685B}) by fitting a two-temperature model to spectra of the 9 radial bins, finding that the plasma is well described by a single temperature model; thus, we conclude that the abundances are not being underestimated. We also exclude that the inverse iron bias (see e.g., \citealt{2010ecsa.conf...94G}) is responsible for the higher abundance at the cavities edges, as this effect is typical of clusters with temperatures of 2 - 4 keV - while RBS~797 has temperatures that do not fall below 4 keV (see \S\ref{subsec:radia}). We thus conclude that the \textit{Chandra} data hint at a slight excess in metallicity between 20 - 50 kpc  with respect to the central gas, which may be due to the AGN pushing enriched gas outwards while inflating its radio lobes.

\section{The central X-ray point source} \label{subsec:point}
At the center of the \textit{Chandra} image of RBS~797 there is a bright X-ray point source that coincides with the radio core of the AGN. Given its location, this point source likely represent the non-thermal X-ray emission from the nucleus of the radio galaxy. Indeed, the study of the previous \textit{Chandra} exposures found that the X-ray source is well described by an absorbed power-law \citep{2011ApJ...732...71C}. To measure its properties with our deeper X-ray observations, we extracted the spectrum of a circle with radius 1.5$''$ ($\sim$7 kpc, which encloses 90\% of the encircled energy fraction) centered on the source. The background spectrum was extracted from an annulus extending between 2$''$ to 6$''$ from the center. The resulting source spectrum has roughly 5000 net counts in the 0.5 - 7 keV band. 
\\ \indent As also observed by \citet{2011ApJ...732...71C}, we found that the spectrum has the typical appearance of a slightly absorbed power-law. There are no indications for the presence of residual thermal components possibly related to the central ICM, suggesting that the plasma properties between 2$''$ - 6$''$ (the local background extraction region) and those within 1.5$''$ from the center are similar (see also the rather flat temperature and metallicity profiles of Perseus in this radial range, e.g., \citealt{2002MNRAS.337...71S, 2006MNRAS.366..417F}). Using a Galactic-absorbed power-law model (\texttt{tbabs$\ast$po}, without intrinsic absorption) to describe the spectrum results in a rather poor fit: the unphysically flat power-law index ($\Gamma = 0.31\pm0.04$) and the $\chi^{2}/D.o.f. = 548.6/373 = 1.47$ indicate that an additional component is likely required. Indeed, adding an intrinsic absorber (\texttt{ztbabs}) at the cluster redshift of 0.354 resulted in a good fit: we find a modest intrinsic column density of $N_{H}^{int} = (7.49\pm0.82)\times10^{22}$ cm$^{-2}$ and power-law index $\Gamma = 1.70\pm0.12$, for a $\chi^{2}/D.o.f. = 385.9/372 = 1.03$. The improvement is thus statistically significant (F-test value of 156.8 and null-hypothesis probability $2\times10^{-30}$). \\ \indent The value of $N_{H}^{int}$ is consistent with previous estimates \citep{2011ApJ...732...71C}, and suggests that the AGN is not heavily obscured. The steep power-law index is typical of radio galaxies in BCGs; converting the power-law index to spectral index ($\alpha_{X} = \Gamma - 1 = 0.7$) we find a good agreement with the spectral index measured at GHz radio frequencies ($\alpha_{R}\sim$0.9, \citealt{2006A&A...448..853G}). The 2 - 10 keV power-law luminosity is $L_{\text{2-10\,keV}}~=~(1.37\pm0.04)~\times~10^{44}$~erg~s$^{-1}$, while the bolometric power-law luminosity is $L^{X}_{\text{bol}}~=~(4.18\pm0.02)\times10^{44}$ erg s$^{-1}$. The 2 - 10 keV luminosity and the 5 GHz luminosity of the unresolved radio core of the AGN ($5.4\times10^{40}$ erg s$^{-1}$, \citealt{2006A&A...448..853G}) can be combined to estimate the mass of the supermassive black hole in RBS~797 using the fundamental black hole plane \citep{2019ApJ...871...80G}, obtaining M$_{BH}\sim1.4\times10^{9}$~M$_{\odot}$. This value is in good agreement with the estimate of M$_{BH}\sim1.5\times10^{9}$ M$_{\odot}$ based on the central velocity dispersion \citep{2011ApJ...732...71C}.

\section{Discussion} \label{sec:disc}
The three weak shocks propagating in the ICM indicate that AGN feedback in RBS~797 has perturbed the environment. The discovery of shocks in ellipticals, groups and clusters has been limited mainly because of the large exposures typically needed to detect such features, and only a few tens of objects with weak shocks due to AGN activity are known. The number of known systems with multiple shock fronts is even lower (M87, \citealt{2017ApJ...844..122F}, Abell 2052, \citealt{Blanton_2009}, and NGC5813, \citealt{2015ApJ...805..112R}), making RBS~797 the fourth object known to have more than one shock and the farthest in redshift. Coupled with the presence of multiple X-ray cavities in its cool core \citep{2001A&A...376L..27S,2011ApJ...732...71C,2012ApJ...753...47D,2021ApJ...923L..25U}, this makes RBS~797 one of the rare windows to investigate feedback history and ICM heating in clusters.
\subsection{The episodic AGN feedback in RBS~797}\label{subsec:epfeed}
\noindent The analysis of the weak shocks in RBS~797 (\S\ref{sec:result}) revealed that the power released by the AGN in shocks has not changed drastically in between outbursts, injecting roughly $10^{61}$ erg about every 25 Myr. 
\begin{figure}[ht]
	\centering
	\gridline{\fig{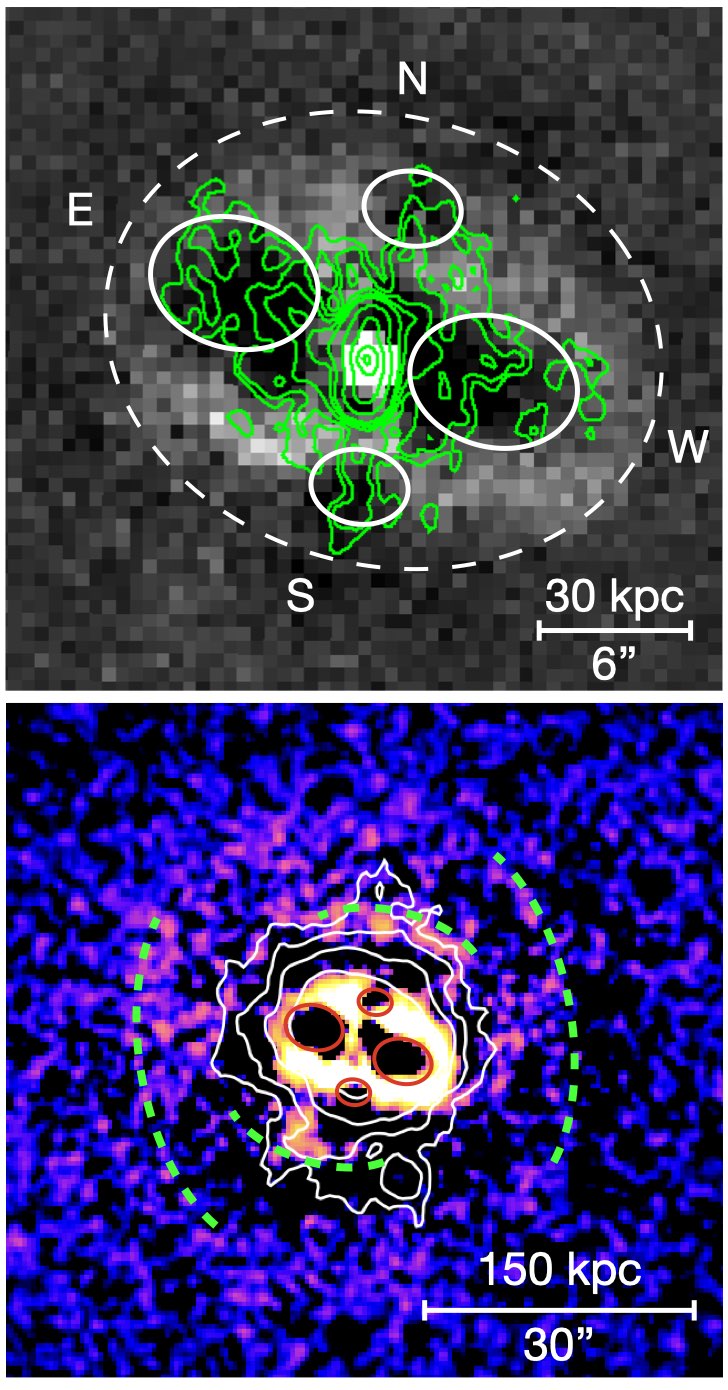}{0.95\linewidth}{}}
	\caption{Residual \textit{Chandra} images of RBS~797. \textit{Upper panel:} zoom on the inner $\sim$60 kpc. The four white ellipses and corresponding labels indicate the shape and name of the X-ray cavities. The dashed white ellipse traces the inner coocon shock. Green contours at 3 GHz from the new JVLA observations (see \S\ref{subsec:radio}; Ubertosi et al., in prep.) are overlaid. The resolution is 0.9$''\times$0.8$''$, the rms noise is 5 $\mu$Jy/beam, and the contours are drawn at (6, 8, 10, 12, 16, 48, 96, 256, 360)$\times$rms. \textit{Lower panel:} residual image of the inner $\sim$300 kpc. Green dashed arcs mark the position of the middle and the outer shock fronts. White contours at 1.4 GHz from the VLA (3$''$ resolution, see \S\ref{subsec:radio} for details) show the diffuse radio emission surrounding the AGN (see \citealt{2012ApJ...753...47D}).}
	\label{fig:rbsmini}
\end{figure}
We now aim to perform a similar evaluation for the four X-ray cavities at the center of RBS~797. The first results from the deep exposure indicated that the two pairs of cavities on perpendicular axes have similar ages (around $\sim$30 Myr, \citealt{2021ApJ...923L..25U}). The preliminary analysis of our new JVLA data (Ubertosi et al. in prep.) further supports the existence of the four cavities, as the 3 GHz radio contours (Fig. \ref{fig:rbsmini}, top panel) are found for the first time to nicely fill also the newly discovered N-S cavities.
\\ \indent To determine the energetics of the four X-ray cavities (whose shape is shown with white ellipses in Fig. \ref{fig:rbsmini}), we consider that $H =  [\gamma/(\gamma-1)]pV = 4pV$ (assuming $\gamma=4/3$) is the enthalpy of a cavity (where $p$ is the pressure of the surrounding ICM and $V$ is the cavity volume), which is also the minimum total energy required to inflate the cavity (e.g., \citealt{2004ApJ...607..800B}). From the pressure profile shown in Fig. \ref{fig:therm} we find an ICM pressure at the average distance of the cavities from the center ($\sim$27 kpc) of $p = 9.7\pm0.8 \times10^{-10}$ erg cm$^{-3}$. The volumes have been determined from the sizes reported in table 1 of \citet{2021ApJ...923L..25U}, by assuming that the cavities are prolate ellipsoids and including a 10\% relative uncertainty. Results are reported in Tab. \ref{tab:cavage}. Then, to compute the cavity power $P_{\text{cav}} = E_{cav}/t_{\text{age}}$, we considered the age estimates we derived in \citet{2021ApJ...923L..25U}. To account for the different methods used to determine the age of each cavity (sound speed, buoyancy, refill and expansion timescale, see e.g., \citealt{2012NJPh...14e5023M}), we computed the mean of the four values for each cavity (third column in Tab. \ref{tab:cavage}), assuming the scatter as an estimate of our uncertainty (which is larger than statistical errors). The similar ages of the bubbles are compatible with the E-W and N-S X-ray cavities being nearly coeval \citep{2021ApJ...923L..25U}.
\begin{table}
	\centering
	\caption{Energetics of the four X-ray cavities in RBS 797.}
	\label{tab:cavage}
		\begin{tabular}{l|ccc}
			\hline
			 & $E_{cav}$ & $t_{\text{age}}$ & $P_{\text{cav}}$  \\
			& [$10^{60}$ erg] & [Myr] & [$10^{45}$ erg s$^{-1}$] \\
			\hline
			
			\rule{0pt}{4ex} Cavity E& 1.4$\pm$0.2 & 30.9$\pm$15.6 & 1.4$\pm$0.9 \\

			\rule{0pt}{4ex} Cavity W & 1.4$\pm$0.2 & 29.5$\pm$16.2 & 1.5$\pm$0.9  \\
			
			\hline
			
			\rule{0pt}{4ex} Cavity N & 0.29$\pm$0.04 & 29.2$\pm$13.3 & 0.32$\pm$0.17  \\

			\rule{0pt}{4ex} Cavity S & 0.29$\pm$0.04 & 27.7$\pm$12.9 & 0.34$\pm$0.18 \\
		
			\hline
		\end{tabular}
		\tablecomments{(1) Cavity name (see figure 1 in \citealt{2021ApJ...923L..25U}; see also Fig. \ref{fig:rbsmini}); (2) Energy of the cavity, computed as $E_{cav}=4pV$ (see text for details); (3) Age of the X-ray cavity, computed as the average of the sound cross, buoyancy, refill and expansion timescales reported in table 1 of \citealt{2021ApJ...923L..25U}; (4) Cavity power $P_{\text{cav}}=E_{cav}/t_{\text{age}}$.}
\end{table}
\\ \indent Overall, the difference in power between the E-W ($\sim3\times10^{45}$ erg s$^{-1}$) and the N-S ($\sim6\times10^{44}$ erg s$^{-1}$) cavity pairs reflects the difference in energy (a factor of 5), which in turn is caused by the E-W cavities being larger than the N-S ones. In \citet{2021ApJ...923L..25U} we argued that the nearly coeval cavities are consistent with both a rapid reorientation of the AGN jets and the presence of binary AGNs in the BCG. Both scenarios may account for the different power: either the AGN has reoriented its jets \textit{and} reduced/increased its mechanical power (see also e.g., MS~0735.6+7421, \citealt{2014MNRAS.442.3192V}), or the X-ray cavities might have been excavated by two AGNs with different mechanical power. 
\\ \indent  By summing the values of each cavity (Tab. \ref{tab:cavage}, second column for energy and fourth column for power), we deduce that the total energy stored in the four X-ray cavities is $E_{cav}^{tot}=3.3\pm0.5\times10^{60}$ erg, while the total power is $P_{\text{cav}}^{tot}=3.6\pm2.2\times10^{45}$ erg s$^{-1}$. By comparing these values with the total energy carried by shocks, we find that $E_{\text{sh}}^{tot}/E_{cav}^{tot} \sim 18$. This value is slightly higher than those of other galaxy clusters and groups with shocks and X-ray cavities, where the $E_{\text{sh}}/E_{cav}$ ratio ranges between 0.1 - 10 \citep{2019MNRAS.484.3376L}. It is worth mentioning, however, that if we are missing older X-ray cavities associated with the middle and outer shocks, then the ratio obtained for RBS~797 might be overestimated. 
It is useful to compare RBS~797 with the galaxy group NGC~5813, where \citet{2011ApJ...726...86R,2015ApJ...805..112R} found three collinear pairs of X-ray cavities, each associated with an elliptical shock front. First, it is interesting to note that both NGC~5813 and RBS~797 have an outburst repetition rate of 1 shock every $\approx2\times10^{7}$ yr. Thus, at least for these two examples, we observe that the repetition interval of low mass systems is similar to that of high mass systems. Nonetheless, while in NGC~5813 the alignment of the outbursts enables a clear association of each X-ray cavity system with its shock, the misaligned cycles of AGN activity in RBS~797 determined a complex ICM geometry, which in turn prevents us from clearly linking X-ray cavities and shocks.
\\ \indent In this context, we note the good agreement between the age of the E-W/N-S cavities ($\sim$28 - 31 Myr) and the age of the inner edge ($t_{\text{age}}^{S_{\text{in}}}=33.4\pm1.3$ Myr). This similarity in timescales further strengthens the association between the X-ray cavities and the inner shock already evident in the morphology. Furthermore, we may speculate that the inner shock, being stronger in the E-W direction (see \S\ref{subsec:wshocks}), has been driven by the same jet episode that inflated the E-W X-ray cavities. This hypothesis could suggest that the N-S cavities are younger than the E-W ones (by a factor smaller than our uncertainties, that point to the cavities being nearly coeval; see Tab. \ref{tab:cavage} and the discussion in \citealt{2021ApJ...923L..25U}), since otherwise the inner cocoon shock would have disrupted any previous N-S small bubble (see e.g., the example of M87 in \citealt{2001ApJ...554..261C}).  However, projection effects may hide the true 3D distribution of the cavities within the cocoon shock, which in turn prevents us from drawing firm conclusions on which jet activity episode is responsible for the inner shock front. Moreover, even though the N-S cavity power indicates a weaker outburst with respect to the E-W one, the N-S jet activity may also have driven weak shock waves that we are unable to detect. Indeed, any AGN jet episode can drive weak shocks, regardless of the jet power (see the simulations of \citealt{2021MNRAS.506..488B}), with shocks driven by weaker jets quickly broadening into sound waves (which are difficult to detect even in local, deeply observed clusters, see e.g., \citealt{2008MNRAS.391.1749G}).
\\ \indent Regarding the middle and outer shocks, the detection of other, external X-ray cavities is likely prevented by the increasing difficulty in detecting cavities at large distances from the center (e.g., \citealt{2009AIPC.1201..301B}) and by the complicated geometry of the cluster. For instance, older cavities produced $\sim$80 Myrs ago (when the outer shock was launched) may have been overridden and distorted by the passage of the middle shock $\sim$54 Myr ago (see also \citealt{2014ApJ...782L..19B}). In this respect, we notice that RBS~797 shows diffuse radio emission within its cool core, with an extent of $\approx$100 kpc and slightly elongated to N-S. The diffuse radio source has been classified as a radio mini-halo \citep{2006A&A...448..853G,2012ApJ...753...47D}. Without entering a detailed discussion, we may speculate a connection between the mechanical power driven by the AGN activity in the ICM and the diffuse radio emission. Shocks may (re)-accelerate relativistic seeds from AGN outbursts via several mechanisms including compression and turbulence (generated downstream by shocks themselves), and advect the plasma on 100 kpc scales (e.g., \citealt{2014IJMPD..2330007B}). By over-plotting the 1.4 GHz VLA radio contours on the \textit{Chandra} residual image in the lower panel of Fig. \ref{fig:rbsmini}, we notice that the N-S radio protrusions are roughly parallel to the axis of outer shock, while the roundish structure outside the cavities seems to be caged within the middle shock. It may thus be possible that we are observing the remnant of what once were distinct AGN outbursts in several directions: due to the passage of multiple shocks the radio emission may have been distorted until a rather amorphous and nearly circular shape has formed.

\subsection{Are heating and cooling balanced in RBS~797?}\label{subsec:heatcool}
\noindent To test the ICM-AGN feedback cycle paradigm, we aim at comparing the shock and cavity powers with the amount of radiative losses in the X-ray band. Several studies in the literature have considered that the gas bolometric X-ray luminosity within the cooling radius of cool core clusters can be considered as a proxy for the magnitude of ICM cooling and flowing to the center (e.g., \citealt{1984Natur.310..733F,2004ApJ...607..800B,2006ApJ...652..216R}). \\ \indent We determined that RBS~797 has a cool core (where $t_{\text{cool}}\leq7.7$ Gyr) with a cooling radius of $109$ kpc (see \S\ref{subsec:radia}). We thus extracted the spectrum of the gas within the cooling radius (excluding the inner 1.5$''$) and of the region between 109 kpc - 500 kpc to allow for deprojection. 
The spectra were fitted with a deprojected thermal model (\texttt{projct$\ast$tbabs$\ast$apec}), obtaining the following results:
\begin{itemize}[noitemsep,leftmargin=*]
    \item \textit{For $r \leq r_{\text{cool}}$} the ICM has a temperature $kT = 5.55^{+0.03}_{-0.05}$ keV, an abundance $Z = 0.56\pm0.02$ Z$_{\odot}$ and radiates a bolometric luminosity of $L_{bol}^{X}\equiv L_{\text{cool}} = (2.34\pm0.01)\times10^{45}$ erg s$^{-1}$. 
    \item \textit{For $r > r_{\text{cool}}$} we measure a temperature $kT = 8.28\pm0.12$ keV, an abundance $Z = 0.40\pm0.03$ Z$_{\odot}$ and a bolometric luminosity of $L_{bol}^{X}= (1.45\pm0.02)\times10^{45}$ erg s$^{-1}$. The $\chi^{2}/D.o.f.$ is 5896/6012 ($=0.98$). 
\end{itemize}
We also tested the inclusion of an isobaric cooling component (\texttt{mkcflow}) to constrain the spectroscopic mass deposition rate, i.e. the rate at which gas is actually cooling to lower temperatures, by fitting a \texttt{tbabs$\ast$(apec+mkcflow)} model to the spectrum of the ICM for $r\leq r_{\text{cool}}$. The high temperature and abundance of the \texttt{mkcflow} component were tied to the values of the \texttt{apec}, while the low temperature was fixed at 0.1 keV. We found that we can only place an upper limit on the mass deposition rate of $\dot{M}_{\text{cool}} \leq 37 \,\text{M}_{\odot}\,\text{yr}^{-1}$ (at 99\% confidence). By reducing the size of the extraction region to $r\leq30$ kpc (where the entropy is smaller than 30 keV cm$^{-2}$, see \S\ref{subsec:radia}), we still measure an upper limit on the mass deposition rate of $\dot{M}_{\text{cool}}\leq 40 \,\text{M}_{\odot}\,\text{yr}^{-1}$. We note that \citet{2012ApJ...753...47D} obtained a $\dot{M}_{\text{cool}} = 231^{+316}_{-227}\,\text{M}_{\odot}\,\text{yr}^{-1}$ from the previous \textit{Chandra} observations. While this estimate would suggest a much higher mass deposition rate, the associated large uncertainties are fully consistent with our more stringent upper limit.
\setlength{\tabcolsep}{6pt}
\begin{table*}[ht]
	\centering
	\caption{Systems with X-ray cavities and weak shocks, ordered by decreasing X-ray cooling luminosity.}
	\label{tab:heatcool}
		\begin{tabular}{l|c|c|c|c|c|c|c}
			\hline
			Name & $L_{\text{cool}}$ & $P_{\text{sh}}$ & $P_{\text{cav}}$ & $P_{\text{tot}}$ & FR-class & $K_{0}$ & Reference \\
			 & [$10^{43}$ erg s$^{-1}$] & [$10^{43}$ erg s$^{-1}$] & [$10^{43}$ erg s$^{-1}$] & [$10^{43}$ erg s$^{-1}$] & & [keV cm$^{2}$]& \\
			\hline
			
			RBS~797      & 234   & 2900 &  360 & 3360 & FR~I & 20.0$\pm$2.4 & This work \\
			
			Perseus      &  67   &  120 &   32 &  152 & FR~I & 19.4$\pm$0.3 & [1,2] \\
			
			MS0735.6+7421 &  26   & 1100 & 1700 & 2800 & FR~II& 16.0$\pm$3.2 & [3,4] \\
			
			Hydra~A      &  25   &  200 &  210 &  410 & FR~I & 13.3$\pm$0.7 & [1,5,6] \\
			
			Cygnus~A     &  29   & 1000 &  210 & 1210 & FR~II& 23.6$\pm$0.9 & [1,7,8] \\
			
			A~2052       &   8.4 &  1.0 &  3.2 &  4.2 & FR~I & 9.5$\pm$0.7 & [1,9] \\
			
			3C~444       &   8.0 & 2900 &  61 & 2961 & FR~II& - & [10,11] \\
			
			Centaurus~A  &   3.0 &  1.2 & 0.74 & 1.94 & FR~I & 2.3$\pm$0.1 & [1,12,13] \\
			
			3C~310       &   1.7 &  190 &  130 &  320 & FR~II& - & [14] \\
			
			M~87         &  0.98 &  2.4 &  1.0 &  3.4 & FR~I & 3.5$\pm$0.1 & [1,15,16,17,18] \\
			
			HCG~62       &  0.15 &  4.0 & 0.38 &  4.4 & FR~I & 3.4$\pm$0.1 & [19] \\
			
			NGC~5813     & 0.055 &  3.5 & 0.18 &  3.7 & FR~I & 1.4$\pm$0.2 & [20,21] \\
			
			3C~88        & 0.055 &   10 &  2.0 &   12 & FR~II& 7.3$\pm$1.7$^{(*)}$ & [22] \\
			
			NGC~4636     & 0.030 & 0.16 & 0.11 & 0.27 & FR~I & 1.4$\pm$0.1 & [23,24,25]  \\
			
			NGC~4552     &0.0025 & 0.33 &0.015 & 0.35 & FR~I & - & [26] \\
			\hline
		\end{tabular}
		\tablecomments{(1) Name; (2) cooling luminosity (3) total shock power; (4) total cavity power; (5) total heating power, defined as $P_{\text{tot}} = P_{\text{cav}} + P_{\text{sh}}$; (6) FR class of the central radio galaxy (based on radio morphology and/or radio power, see literature references in column 8); (7) central entropy from \citet{2009ApJS..182...12C}, defined as the excess entropy above the best-fitting power law found at larger radii; (8) Literature references for the values reported in columns 2 - 6: [1] \citet{2004ApJ...607..800B}, [2] \citet{2008MNRAS.386..278G}, [3] \citet{2014MNRAS.442.3192V}, [4] \citet{Biava2021}, [5] \citet{2005ApJ...628..629N}, [6] \citet{2007ApJ...659.1153W}, [7] \citet{2012MNRAS.427.3468B}, [8] \citet{2018ApJ...855...71S}, [9] \citet{Blanton_2009}, [10] \citet{Croston_2011}, [11] \citet{2017MNRAS.466.2054V}, [12] \citet{2006ApJ...652..216R}, [13] \citet{10.1111/j.1365-2966.2009.14715.x}, [14] \citet{Kraft_2012}, [15] \citet{2001ApJ...554..261C}, [16] \citet{2005ApJ...635..894F}, [17] \citet{2007ApJ...665.1057F}, [18] \citet{2017ApJ...844..122F}, [19] \citet{2010ApJ...714..758G}, [20] \citet{2011ApJ...726...86R}, [21] \citet{2015ApJ...805..112R}, [22] \citet{2019MNRAS.484.3376L}, [23] \citet{2002ApJ...567L.115J}, [24] \citet{2005ApJ...624L..77O}, [25] \citet{2009ApJ...707.1034B}, [26] \citet{2006ApJ...648..947M}.
		\\$^{(*)}$: The value is taken from \citet{2019MNRAS.484.3376L}.
		}
\end{table*}
\\ \indent  Altogether, we find that cavity power alone ($P_{\text{cav}}^{tot} = 3.6\times10^{45}$ erg s$^{-1}$) can balance radiative losses within the cool core of RBS~797. Including the energy injected in shocks, the total heating power of the AGN in RBS~797 is of roughly $3.4\times10^{46}$ erg s$^{-1}$, which exceeds the cooling luminosity, $L_{\text{cool}}=(2.34\pm0.01)\times10^{45}$ erg s$^{-1}$ by a factor $\sim14$.  
Additionally, we note that the outer shock, which is located at $\sim130$ kpc from the center, lies outside the cool core ($r_{\text{cool}}\sim109$ kpc). Thus, the AGN in RBS~797 may not only be able to match radiative losses within the cool core, but also to heat the gas at larger distances from the center. Similar results were obtained in the cases of Hydra~A (e.g. \citealt{2005ApJ...628..629N,2007ApJ...659.1153W}), and of MS~0735.6+7421 (e.g., \citealt{2005Natur.433...45M,2007ApJ...660.1118G}), where however giant cavities (with radius exceeding 100 kpc) and large-scale shock fronts (at distances of more than 200 kpc from the center) were found. The fact that in RBS~797 smaller shock fronts (and the possibly associated undetected cavities) lie outside $r_{\text{cool}}$ may imply that heating beyond the cool core is more common than previously thought (i.e. there may be undetectable cavities and shock fronts or sound waves at $r\geq r_{\text{cool}}$ in many other systems).
\\ \indent We can also test the local balance of shock heating and radiative cooling. Following the strategy applied by \citet{2011ApJ...726...86R,2015ApJ...805..112R} to NGC~5813, we consider that the fractional effective heat input from one shock is $\Delta \text{ln}(p/\rho^{\gamma})$. Considering the Mach numbers of the shock fronts (see Tab. \ref{tab:specfronts}), we find that the change in $\text{ln}(p/\rho^{\gamma})$ across the inner, middle, and outer shocks is $\sim$1\%, $\sim$0.4\% and $\sim$0.5\%, respectively. Taking the reciprocal of these values, we find that $\sim$100, $\sim$250, and $\sim$200  outbursts are needed per local cooling time to completely offset cooling with shock heating at the location of the inner, middle and outer shock edges, respectively. The cooling time outside the edges is 2 Gyr (inner shock), 5 Gyr (middle shock) and 9 Gyr (outer shock). Assuming an outburst interval of 25 Myr (see \S\ref{subsec:epfeed}) gives 80 shocks (inner shock), 200 shocks (medium shock) and 360 shocks (outer shock) per local cooling time. Thus, we find that in RBS~797 there is agreement between the number of shocks required to offset radiative cooling and those expected per local cooling time (as found by \citealt{2011ApJ...726...86R,2015ApJ...805..112R} for the galaxy group NGC~5813). 

\subsection{Shock and cavity heating of hot atmospheres}\label{subsec:sampleshock}
\begin{figure}[ht!]
	\centering
	\gridline{\fig{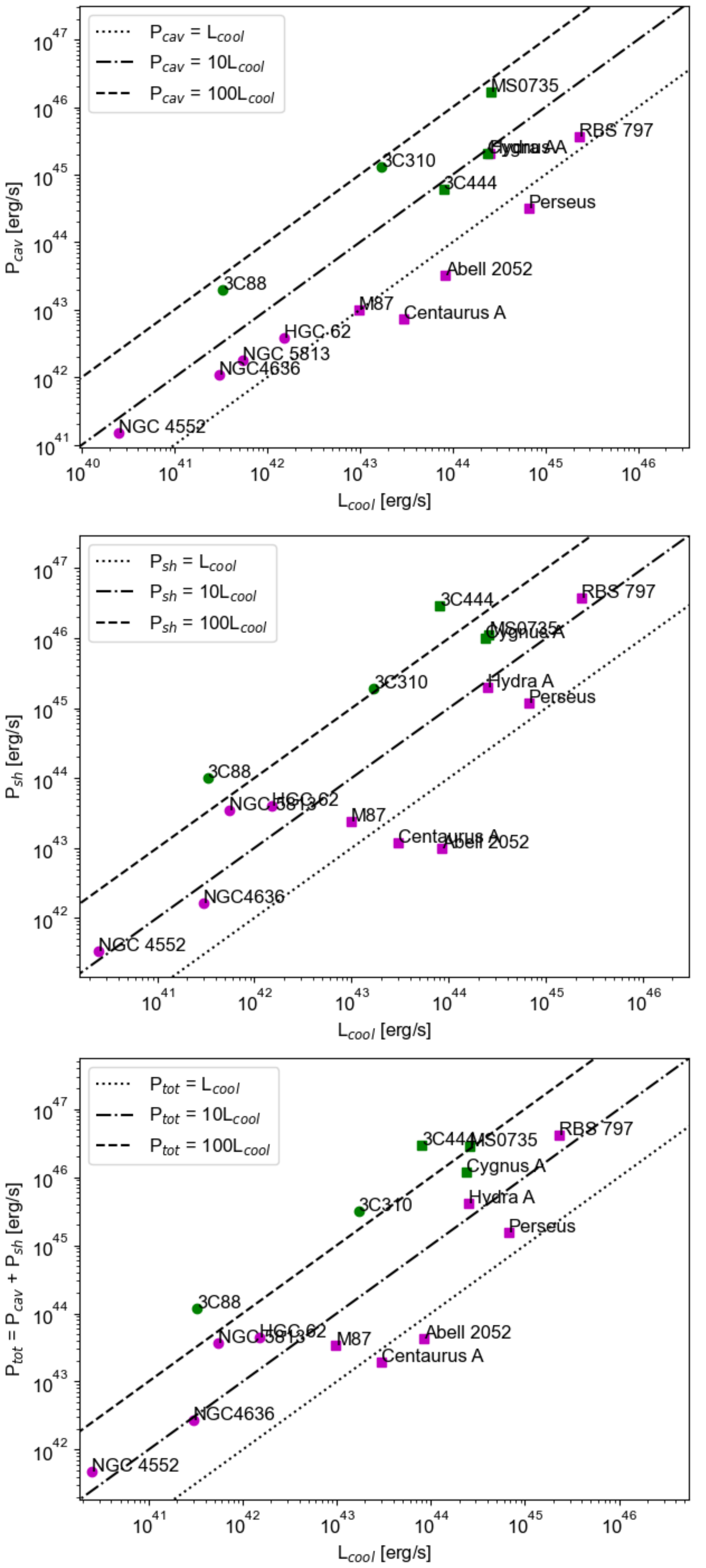}{\linewidth}{}}
	\vspace{-0.7cm}
	\caption{Mechanical cavity power (\textit{upper panel}), shock power (\textit{middle panel}) and total heating power (\textit{lower panel}) versus cooling luminosity for the 15 systems listed in Tab. \ref{tab:heatcool}. Overplotted are lines of $P/L_{\text{cool}} = 1,\,10,\,100$. Dots are for elliptical galaxies and galaxy groups, while squares are for galaxy clusters. The color of the points indicates the FR type of central radio galaxy: FR~Is are in magenta, while FR~IIs are in green.}
	\label{fig:shockcool}
\end{figure}
The total AGN mechanical power in RBS~797 (which is larger than $L_{\text{cool}}$ by a factor $\sim$14) may seem extreme with respect to systems in typical samples of galaxy clusters, galaxy groups and elliptical galaxies, where the average ratio between mechanical power and cooling luminosity is close to unity (e.g., \citealt{2004ApJ...607..800B,2006ApJ...652..216R}). However, such samples are mostly composed of systems where only X-ray cavities have been found, whereas in RBS~797 we also consider the energy injected by shocks, that dominates the AGN power. Indeed, by considering only the X-ray cavities, we find a ratio $P_{\text{cav}}/L_{\text{cool}}\sim1$. Thus, a more consistent comparison should be made with other systems with detections of both X-ray cavities and weak shocks.
\\ \indent  \citet{2019MNRAS.484.3376L} assembled a compilation of 13 galaxy clusters, groups and ellipticals in which X-ray cavities and shocks have been detected, and for which Mach number, shock energy and cavity energies are available. As we are interested in the comparison between the shock power, X-ray cavity power, average total (= shock + cavity) heating power and cooling luminosity, first we verified that these 13 systems also have available measurements of such quantities. Additionally, we checked whether new detections had been claimed after the work of \citet{2019MNRAS.484.3376L}, finding no new results. Surface brightness edges have been identified in 3C~220.1 by \citet{2020MNRAS.492.3156L} and in 3C~320 by \citet{2019MNRAS.485.1981V}, however due to the insufficient number of counts a spectral confirmation of their shock front nature was not performed. We also consider here Perseus due to the putative weak shock surrounding its inner X-ray cavities \citep{2008MNRAS.386..278G}. We report in Tab. \ref{tab:heatcool} the properties of the 15 systems (13 from \citealt{2019MNRAS.484.3376L} + Perseus + RBS~797) with corresponding literature references. We find that RBS~797 has the largest cooling luminosity and largest total power among the 15 objects.
\\ \indent In the three panels of Fig. \ref{fig:shockcool} we plot the comparison of $P_{\text{cav}}\,\text{vs.}\,L_{\text{cool}}$, $P_{\text{sh}}\,\text{vs.}\,L_{\text{cool}}$ and $P_{\text{tot}}\,\text{vs.}\,L_{\text{cool}}$ for the objects we selected. The first panel reveals the trend already well tested (e.g., \citealt{2004ApJ...607..800B,2006ApJ...652..216R,2012MNRAS.427.3468B}) between the mechanical power of X-ray cavities and the cooling luminosity of the ambient gas. With a span in $P_{\text{cav}}$ and $L_{\text{cool}}$ of about five orders of magnitude, these 15 systems confirm that larger ICM radiative losses are accounted for by more powerful outbursts. In the second panel we show that also the shock power scales with the X-ray cooling luminosity, suggesting that the whole AGN mechanical output is set in response to the amount of available fuel. We note that the majority of the systems have $P_{\text{cav}}/L_{\text{cool}}\geq1$ and $P_{\text{sh}}/L_{\text{cool}}\geq1$ (dashed lines in the first two panels of Fig. \ref{fig:shockcool}); given that shocks are more likely found for powerful objects, it is unsurprising that our selection criteria identify systems where cooling is largely balanced by mechanical feedback. These trends are consistent with the plot in the third panel, which shows the scaling between the total average heating power $P_{\text{tot}} = P_{\text{cav}} + P_{\text{sh}}$ and the cooling luminosity. 
As mentioned above, while RBS~797 has a high total mechanical power, it may not be in a relative sense an extremely heated object. Indeed, the high cooling luminosity of RBS~797 indicates a ratio $P_{\text{tot}}/L_{\text{cool}}\sim14$, whereas there are systems in the third plot of Fig. \ref{fig:shockcool} that are able to provide total mechanical power that exceeds by two orders of magnitude the radiative losses of their gaseous halos. In particular, we note that the upper area of the $P_{\text{tot}}/L_{\text{cool}}$ space is populated by the 3C sources 3C~88, 3C~310 and 3C~444, by Cygnus~A and by MS~0735.6+7421, which have $P_{\text{tot}}/L_{\text{cool}}$ ratios ranging between 50 (Cygnus~A) to 370 (3C~444). \\ \indent By considering the Fanaroff-Riley class of the radio galaxies in our compilation (see Tab. \ref{tab:heatcool}), we find that the sources named above are all FR~II radio galaxies, while the other systems host FR~I AGNs at their center, characterized by $0.5\leq P_{\text{tot}}/L_{\text{cool}}\leq60$. We note that this distinction holds also for the $P_{\text{cav}}/L_{\text{cool}}$ and $P_{\text{sh}}/L_{\text{cool}}$ comparison (upper and middle panel of Fig. \ref{fig:shockcool}). The two classes seem to overlap at $P_{\text{tot}}/L_{\text{cool}}\sim50-60$, where we find that NGC~5813, a low luminosity FR~I radio galaxy, has a larger ratio than Cygnus~A, the archetypal FR~II radio galaxy. However, the different environments have to be taken into consideration: NGC~5813 is a galaxy group \citep{2011ApJ...726...86R}, while Cygnus~A is at the center of a galaxy cluster (e.g., \citealt{2014A&A...563A.131S}). In this context, previous studies noted that the relative ratio of X-ray cavity heating power to cooling luminosity appears to be 5 times higher in low mass systems than in rich clusters (e.g., \citealt{2012AdAst2012E...6G}). By separating the systems in our compilation between galaxy clusters (squares in Fig. \ref{fig:shockcool}) and galaxy groups and elliptical galaxies (circles in Fig. \ref{fig:shockcool}), we find that FR~IIs have $P_{\text{tot}}/L_{\text{cool}}\geq180$ while FR~Is have $9\leq P_{\text{tot}}/L_{\text{cool}}\leq 66$ in galaxy groups and elliptical galaxies. In galaxy clusters, FR~IIs have $P_{\text{tot}}/L_{\text{cool}}\geq50$ while FR~Is have $0.5\leq P_{\text{tot}}/L_{\text{cool}}\leq 16$. These values are consistent with AGNs in galaxy groups and elliptical galaxies having relatively more power to counter cooling than sources in galaxy clusters - with FR~IIs always providing a more violent feedback.
\\ \indent Overall, it appears that both classes of radio galaxies have mechanical power that correlates with the cooling luminosity of their hot atmospheres, suggesting that FR~Is and FR~IIs can achieve a self-regulated equilibrium between input fuel and output energy. However, FR~IIs seem to be able to sustain a more extreme feedback. In this respect, sources populating the upper space of the $P_{\text{tot}}/L_{\text{cool}}$ plot may be expected to have more efficiently heated the ambient gas and quenched any residual cooling. We can test this scenario by considering that the entropy of the ICM is a good proxy for the cooling efficiency of the gas. \citet{2009ApJS..182...12C} analyzed the entropy profiles of 239 systems (the ACCEPT sample) and measured the central entropy $K_{0}$, that quantifies the typical excess of core entropy above the best-fitting power law found at larger radii. By cross matching the list of Tab. \ref{tab:heatcool} with the ACCEPT sample, we find that 11 out of 15 objects have tabulated values of $K_{0}$ (missing sources are 3C~444, 3C~310, 3C~88 and NGC 4552). We additionally include 3C~88 since \citet{2019MNRAS.484.3376L} determined the core entropy of this group in a similar fashion to that of \citet{2009ApJS..182...12C}. Based on the core entropy values for systems in our list (see Tab. \ref{tab:heatcool}), we find that the hot atmospheres of FR~I radio galaxies have $K_{0}$ in the range 1 - 20 keV cm$^{2}$ (with an average of $9\pm8$ keV cm$^{2}$, where the uncertainty is the dispersion around the mean). For the three FR~IIs matching our criteria, the central entropy is in the range 7 - 24 keV cm$^{2}$ (with an average of $15\pm$8 keV cm$^{2}$). The clear overlap between the two classes indicates that differences of orders of magnitude in feedback output (i.e. the ratio between mechanical power and cooling luminosity) do not cause any over-heating of the central gas. Therefore, we find that the FR~IIs in the plots of Fig. \ref{fig:shockcool}, while providing a stronger feedback, are not dramatically quenching cooling of the surrounding environment. This may suggest that the thermodynamic regulation of the ICM on long timescales (the central cooling time of these systems ranges between hundreds of Myr and $\sim$1 Gyr) is resilient to episodic (the typical outburst ages are of a few tens of Myr), overpowered outburst episodes (for numerical simulations supporting this picture, see e.g., \citealt{2011MNRAS.411..349G,2014MNRAS.441.1270L,2015ApJ...811..108P,2020ApJ...896..114D,2021MNRAS.506..488B}). 
\\ \indent We note that these results are speculative and should be treated with caution. In particular, the values reported in Tab. \ref{tab:heatcool} are subject to different assumptions made in the literature. For instance, cavity power depends on cavity age, which can be estimated with several methods. Moreover, the shock energy and power may be estimated as done in this work (and in e.g. \citealt{2015ApJ...805..112R,2019MNRAS.484.3376L}), or by adopting a point explosion model tailored to the specific object \citep{2005ApJ...628..629N}. In any case, the difference between these varying assumptions is typically within a factor of a few (e.g., \citealt{2004ApJ...607..800B,2011ApJ...726...86R,2015ApJ...805..112R,2019MNRAS.484.3376L}), while the trends we noticed in Fig. \ref{fig:shockcool} refer to differences of at least one order of magnitude. Thus, we argue that the effect of non-uniform assumptions is not dominant. On the other hand, the systems we considered are probably not representative of the population of galaxy clusters, galaxy groups and elliptical galaxies. More likely, they represent the tip of the iceberg of an underlying larger population of objects with X-ray cavities \textit{and} weak shocks launched by the central AGN which have not been detected due to limited statistics. As such, we do not provide correlations of $P_{\text{cav}}\,\text{vs.}\,L_{\text{cool}}$, $P_{\text{sh}}\,\text{vs.}\,L_{\text{cool}}$ or $P_{\text{tot}}\,\text{vs.}\,L_{\text{cool}}$, limiting ourselves to discussing general trends. Enlarging the sample to allow a deeper, statistically consistent comparison is a future perspective of this work.

\section{Conclusions} \label{sec:conc}
Our results can be summarized as follows:
\begin{enumerate}[noitemsep,leftmargin=*]
    \item We discovered that AGN activity in RBS~797 has driven three nested shock fronts, found at projected distances of 50 kpc, 80 kpc and 130 kpc from the center, with Mach numbers in the range 1.2 - 1.3. We find that the total energy required to drive the shocks in RBS~797 is roughly $6\times10^{61}$ erg. The mechanical power does not change drastically in between the successive activity cycles, with the AGN driving every 20-30 Myr a weak shock with power $P_{\text{sh}}\approx10^{46}$ erg s$^{-1}$. Based on the morphology and timescales of the inner cocoon shock and the E-W X-ray cavities (see also \citealt{2021ApJ...923L..25U}), we suggest that the bubbles and the shock likely originate from the same outburst. Furthermore, we hypothesize that the amorphous shape of the radio source surrounding the X-ray cavities (at 50 kpc - 100 kpc from the center) may be caused by the passage of the middle and outer shocks, that could have overrun and distorted pre-existing radio plasma from previous AGN activity.
    \item The inflation of X-ray cavities has left footprints in the ICM: we found hints for a ring of enhanced metallicity surrounding the bubbles between $\sim$30 - 50 kpc from the center, likely explained in the context of mechanical uplift of central enriched gas. The low entropy ($\leq$30 keV cm$^{2}$) and short cooling time ($t_{\text{cool}}\leq1$ Gyr) ICM is preferentially found between and behind the X-ray cavities, which may indicate that the bubbles are stimulating cooling. In this context, the region with the shortest cooling time ($\sim$440 Myr) and minimum $t_{\text{cool}}/t_{\text{ff}}$ ratio ($\sim$18) is located within 20 kpc south of the center, where filamentary patches of [O~II] nebular emission tracing star-forming gas were recently detected \citep{2022ApJ...940..140C}.
    \item We estimated that within the cooling radius of RBS~797 ($r_{\text{cool}}^{7.7\,\text{Gyr}}=109.3\pm1.0$ kpc), the X-ray emitting gas is radiating its energy at the remarkable rate of $L_{\text{cool}} = (2.34\pm0.01)\times10^{45}$ erg s$^{-1}$. Such radiative losses are overcome by the total mechanical power in RBS~797 (by shocks and X-ray cavities), i.e. $3.4\times10^{46}$ erg s$^{-1}$. Moreover, the distance from the center to the outer shock (135 kpc) is larger than the cooling radius, suggesting that the gas outside $r = r_{\text{cool}}$ is also being heated. 
    \item By evaluating shocks and cavity power ($P_{\text{sh}}$ and $P_{\text{cav}}$) versus cooling ($L_{\text{cool}}$) for RBS~797 and 14 other well known galaxy clusters, galaxy groups and elliptical galaxies with detections of X-ray cavities and weak shocks, we find that the already known scaling of $P_{\text{cav}}\,\text{vs.}\,L_{\text{cool}}$ exists also for $P_{\text{sh}} \,\text{vs.}\, L_{\text{cool}}$ and $P_{\text{tot}} [= P_{\text{sh}} + P_{\text{cav}}] \,\text{vs.}\, L_{\text{cool}}$. Additionally, while RBS~797 has the largest mechanical power and cooling luminosity, it is not the object with the most extreme feedback. In particular, we noted that systems with FR~I radio galaxies at their center (as RBS~797) seem to have ratios $P_{\text{tot}}/L_{\text{cool}}$ within a few tens, while systems with central FR~II AGNs have total mechanical power that can exceed by more than two orders of magnitude the cooling luminosity of the surrounding atmosphere. This difference becomes more evident when massive systems (clusters) are distinguished from smaller systems (groups and elliptical galaxies). Nevertheless, the central entropy of systems hosting FR~Is is comparable to that of systems hosting FR~IIs, suggesting that differences of orders of magnitude in feedback output do not over-heat the central gas.
\end{enumerate}
\acknowledgments
We thank the anonimous reviewer for their useful suggestions, that improved our work. We acknowledge financial contribution from the agreement ASI-INAF n.2017-14-H.0 (PI Moretti). Support for this work was provided to MM and MC by the National Aeronautics and Space Administration through Chandra Award Number GO0-21114A issued by the Chandra X-ray Center, which is operated by the Smithsonian Astrophysical Observatory for and on behalf of the National Aeronautics Space Administration under contract NAS8-03060. Additional support was provided to MM and MC from the Space Telescope Science Institute, which is operated by the Association of Universities for Research in Astronomy, Inc., under NASA contract NAS 5–26555. This support is specifically associated with the program HST-GO-16001.002-A. AI acknowledges the European Research Council (ERC) under the European Union's Horizon 2020 research and innovation programme (grant agreement No. 833824). This research has made use of data obtained from the Chandra Data Archive and the Chandra Source Catalog, and software provided by the Chandra X-ray Center (CXC) in the application packages CIAO and Sherpa. The National Radio Astronomy Observatory is a facility of the National Science Foundation operated under cooperative agreement by Associated Universities, Inc.
%

\vspace{5mm}
\facilities{CXO, JVLA.}


\software{\texttt{astropy} \citep{2013A&A...558A..33A,2018AJ....156..123A},  
          \texttt{APLpy} \citep{2012ascl.soft08017R}, \texttt{Numpy} \citep{2011CSE....13b..22V,2020Natur.585..357H}, \texttt{Scipy} \citep{jones2001scipy}, \texttt{CIAO} \citep{2006SPIE.6270E..1VF}, \texttt{XSPEC} \citep{1996ASPC..101...17A}, \texttt{AIPS} \citep{van1996aips}, \texttt{CASA} \citep{2007ASPC..376..127M}. 
          }

\bibliographystyle{aasjournal.bst} 
\bibliography{rbs797.bib} 






\begin{appendix}
\setcounter{table}{0}
\renewcommand{\thetable}{A\arabic{table}}
\renewcommand*{\theHtable}{\thetable}
\setcounter{figure}{0}
\renewcommand{\thefigure}{A\arabic{figure}}
\renewcommand*{\theHfigure}{\thefigure}

\section{Details on the \textit{Chandra} X-ray observations of RBS~797}
In Tab. \ref{OBSIDS} of this section we list the \textit{Chandra} ObsIDs analyzed in this article (see \S\ref{subsec:xray}).
\begin{table}
\centering
\caption{List of the \textit{Chandra} observations of RBS~797 used in this work: (1) number of the observation; (2) \textit{Chandra} instrument; (3) raw exposure time; (4) cleaned exposure time after removal of background flares; (5) Principal Investigator. The last row shows the sum of the uncleaned and cleaned exposures.}
\begin{tabular}{lcccr}
\hline
ObsID  & Instrument & t$_{raw}$ [ks] & t$_{clean}$ [ks] & P.I.      \\ \hline
2202   & ACIS-I     & 11.7          & 9.2                       & Schindler \\ \hline
7902   & ACIS-S     & 38.3          & 37.3                      & McNamara  \\ \hline
22636  & ACIS-S     & 44.5          & 41.4                      & Gitti     \\ \hline
22637  & ACIS-S     & 23.7          & 22.1                      & Gitti     \\ \hline
22638  & ACIS-S     & 19.8          & 18.0                      & Gitti     \\ \hline
22931  & ACIS-S     & 24.7          & 23.4                      & Gitti     \\ \hline
22932  & ACIS-S     & 57.5          & 55.1                      & Gitti     \\ \hline
22933  & ACIS-S     & 24.7          & 22.8                      & Gitti     \\ \hline
22934  & ACIS-S     & 25.7          & 25.0                      & Gitti     \\ \hline
22935  & ACIS-S     & 25.7          & 23.9                      & Gitti     \\ \hline
23332  & ACIS-S     & 43.4          & 39.4                      & Gitti     \\ \hline
24631 & ACIS-S     & 24.7          & 23.3                      & Gitti     \\ \hline
24632  & ACIS-S     & 24.6          & 23.0                      & Gitti     \\ \hline
24852  & ACIS-S     & 43.5          & 39.1                      & Gitti     \\ \hline
24865  & ACIS-S     & 25.7          & 24.4                      & Gitti     \\ \hline
Total  &            & 458.4         & 427.4                     &           \\ \hline
\end{tabular}
\label{OBSIDS}
\end{table}

\section{Details on the morphological and spectral analysis of shocks fronts}
\begin{figure*}[ht]
	\centering
	\gridline{\fig{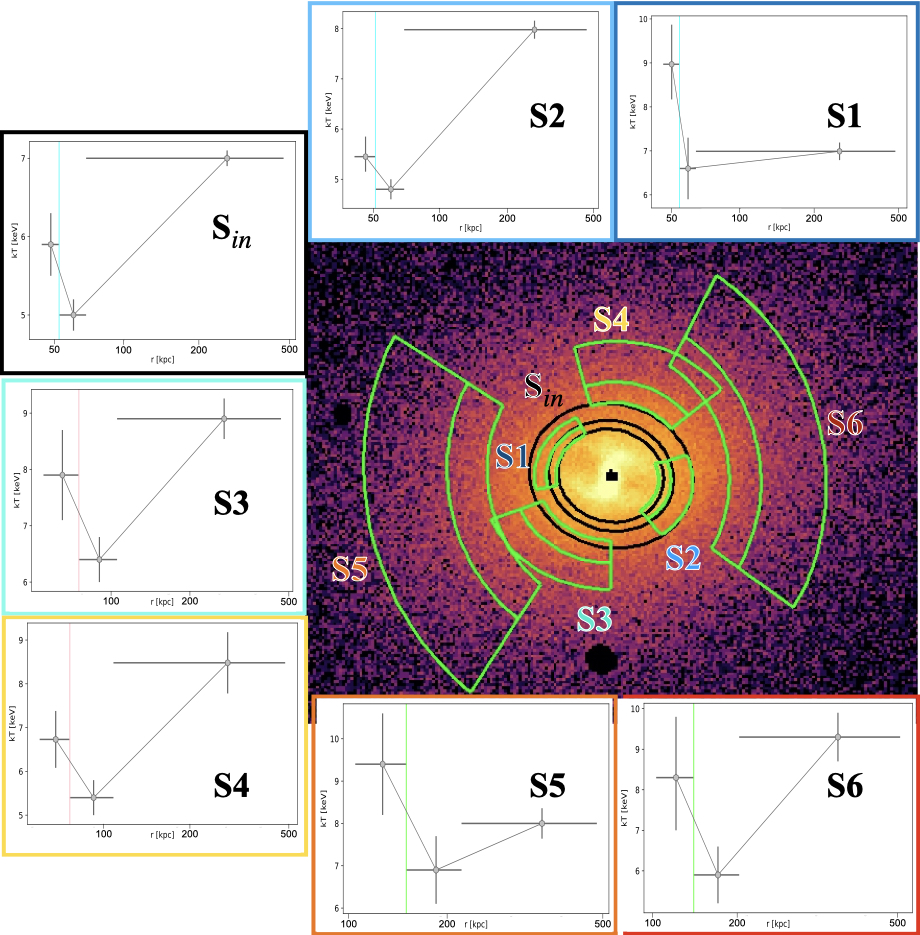}{1\linewidth}{}}
	\caption{\textit{Central panel:} 0.5 - 7 keV \textit{Chandra} image of RBS~797. The green (black) sectors indicate the regions (described in Tab. \ref{tab:fronts}) used to determine the thermodynamic properties across the downstream and upstream sides of the arc-like (cocoon-like) edges. A third region extending to the edge of the \textit{Chandra} image has been used for deprojection (not shown in the figure). Colored labels indicate the name of each edge. \textit{Sub-panels:} Deprojected temperature profile across each front. The vertical colored lines indicate the position of each shock.}
	\label{fig:shockreg}
\end{figure*}

\begin{figure*}[ht]
	\centering
	\gridline{\fig{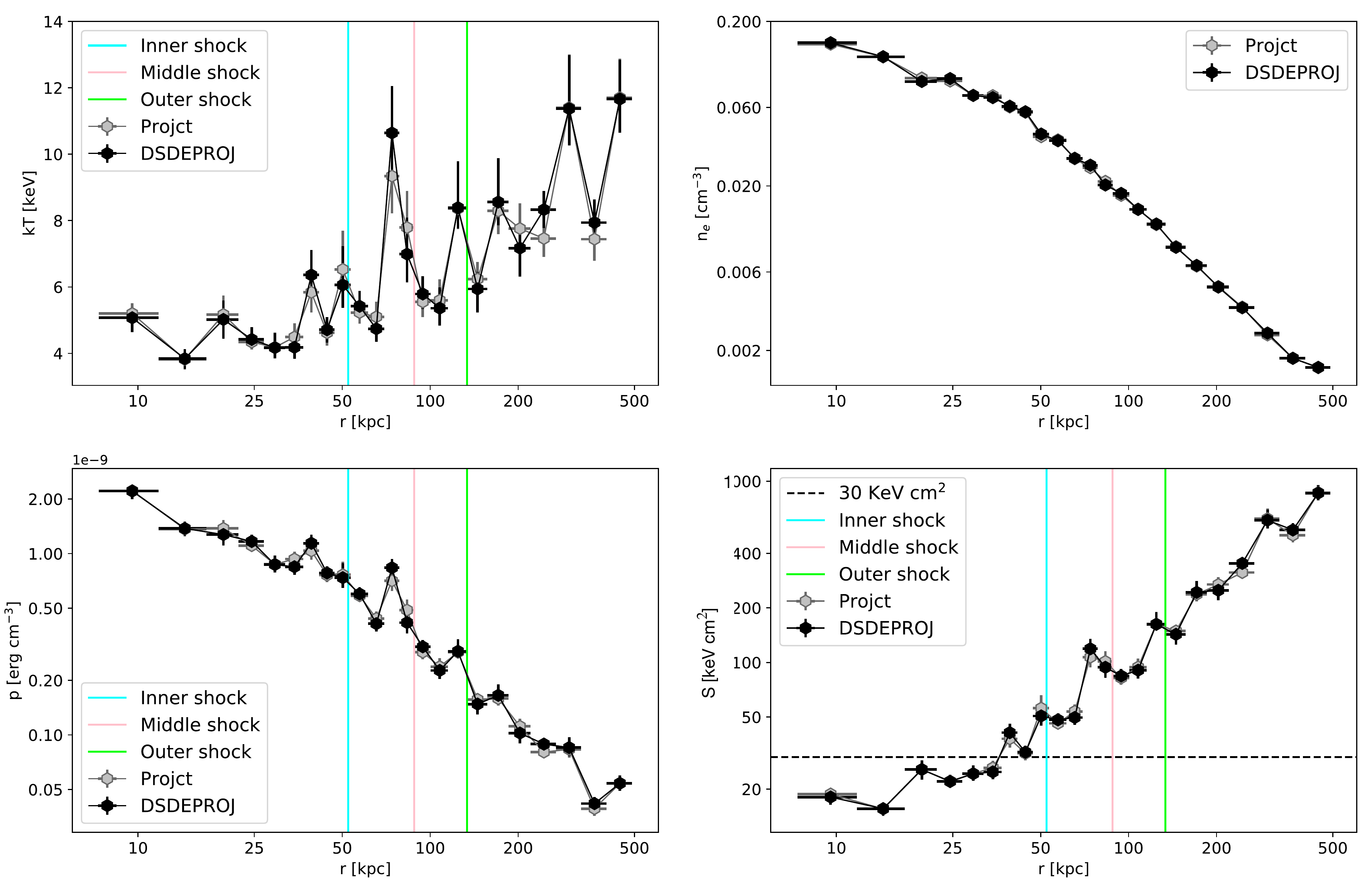}{1\linewidth}{}}
	\caption{Comparison between the radial profiles of ICM thermodynamic quantities in RBS~797 using  \texttt{projct} (gray points) and \texttt{DSDEPROJ} (black points). \textit{Upper left:} deprojected temperature profiles; \textit{Upper right:} electron density profile; \textit{lower left:} pressure profile; \textit{lower right:} entropy profile, with the horizontal dotted line showing the $S\leq30$ keV cm$^{2}$ threshold for the condensation of the ICM into multi-phase gas clouds. In the temperature, pressure, and entropy profiles the colored vertical lines show the distance from the center of the three nested weak shocks.}
	\label{fig:app-pro}
\end{figure*}

In this section we report the detailed results of the analysis of the shock fronts. We show in Tab. \ref{tab:shockbri} the results of fitting surface brightness profiles encompassing each shock fronts with a broken power-law model and a power-law model. Furthermore, we summarize in Tab. \ref{tab:fronts} the geometry of the sectors we used as spectral extraction regions (first three columns), the $\chi^{2}/D.o.f.$ (fourth column) and the best-fit parameters within the downstream and upstream side of the shock. The values of temperature and pressure are deprojected. Fig. \ref{fig:shockreg} shows for each shock front the region used to extract the spectrum of the upstream and downstream side, and the resulting temperature profile. Ultimately, we show in Fig. \ref{fig:app-pro} the comparison between using \texttt{projct} and \texttt{DSDEPROJ} \citep{2007MNRAS.381.1381S,2008MNRAS.390.1207R} to deproject the azimuthally averaged thermodynamic profiles described in \S\ref{subsec:radia}. As it is possible to see, the two methods return consistent deprojected radial profiles. This confirms that the temperature jumps at the location of the shocks are not caused by the ringing effect associated with \texttt{projct}.
\setlength{\tabcolsep}{1.5pt}
\begin{table*}[ht]
	\centering
	\caption{Analysis of the surface brightness profile of the shocks in RBS~797.}
	\label{tab:shockbri}
		\begin{tabular}{c||c|c|c|c|c|c||c|c|c|c||c|c}
			\hline
			
	    &  \multicolumn{6}{c||}{Broken power-law} & \multicolumn{4}{c||}{Power-law} & \multicolumn{2}{c}{F-test} \\
          \cline{2-7} \cline{8-11} \cline{12-13}
		Shock &  $\alpha_{1}$ & $\alpha_{2}$ & $r_{J}$ [$''$] & S$_{0}$ & $J$ &  $\chi^{2}/d.o.f.$ & $\alpha$ & $r_{\text{S}}$ [$''$] & S$_{0}$ & $\chi^{2}/d.o.f.$ &$F$ & $p$ \\
		
			\hline
			
        \rule{0pt}{4ex} S1  & -0.31$\pm$0.08 & 1.35$\pm$0.05 & 11.0$\pm$0.1 & 4.1$\pm$0.2& 1.49$\pm$0.07 & 13.2/7 & 2.14$\pm$0.02& 18.0$\pm$1.0& 0.14$\pm$0.01&195.1/20 &59.3 & 1.2$\times10^{-8}$\\
        
        \rule{0pt}{4ex} S2  & -0.23$\pm$0.09 & 1.63$\pm$0.07 & 10.3$\pm$0.2 & 5.1$\pm$0.6& 1.40$\pm$0.06 & 6.6/4 & 1.99$\pm$0.02& 16.8$\pm$0.6& 0.13$\pm$0.01&176.7/19 &51.6 & 6.0$\times10^{-8}$\\
        
        \rule{0pt}{4ex} S$_{\text{in}}$  & 0.71$\pm$0.05 & 1.36$\pm$0.03 & 10.7$\pm$0.1 & 3.3$\pm$0.2& 1.29$\pm$0.03 & 14.4/8 & 1.67$\pm$0.01& 2.5$\pm$0.1& 4.8$\pm$0.2&353.9/10 &94.0 & 2.8$\times10^{-6}$\\
        
            \hline
            
        \rule{0pt}{4ex} S3  & 1.23$\pm$0.03 & 1.49$\pm$0.04 & 16.2$\pm$0.1 & 0.77$\pm$0.3& 1.29$\pm$0.04 & 25.6/18 & 2.14$\pm$0.01& 18.0$\pm$1.0& 0.14$\pm$0.01&60.6/20 &17.1 & 1.3$\times10^{-5}$\\  
        
        \rule{0pt}{4ex} S4  & 0.26$\pm$0.09 & 1.64$\pm$0.05 & 16.7$\pm$0.2 & 0.94$\pm$0.01& 1.28$\pm$0.05 & 10.1/10 & 2.10$\pm$0.03& 7.2$\pm$0.1& 0.69$\pm$0.07 &127.5/12 &58.1 & 3.1$\times10^{-6}$\\  
        
            \hline
            
        \rule{0pt}{4ex} S5  & 0.32$\pm$0.10 & 1.95$\pm$0.05 & 30.5$\pm$0.4 & 0.23$\pm$0.01& 1.27$\pm$0.04 & 6.1/6 & 2.63$\pm$0.02& 25.2$\pm$0.3& 0.16$\pm$0.01&269.4/8 &130 & 1.1$\times10^{-5}$\\  
        
        \rule{0pt}{4ex} S6  & 0.01$\pm$0.09 & 1.39$\pm$0.04 & 28.9$\pm$0.2 & 0.69$\pm$0.03& 1.37$\pm$0.04 & 24.3/15 & 2.05$\pm$0.02& 19.2$\pm$0.6 & 0.38$\pm$0.02 &193.4/17 &52.2 & 1.8$\times10^{-7}$\\

			\hline
		\end{tabular}
		\tablecomments{For each edge, we report the best-fit parameters using a broken power-law (columns 2 - 7) and a single power-law (columns 8 - 11). The last two columns show the result of an F-test between the two models. The reported $r_{J}$ refers to the distance along the major axis of the ellipse used to describe each edge (see Tab. \ref{tab:fronts} for the geometric details). The normalizations S$_{0}$ are in units of $10^{-2}$ cts s$^{-1}$ arcmin$^{-2}$. Corresponding surface brightness profiles are shown in Fig. \ref{fig:surbr}.}
\end{table*}

\setlength{\tabcolsep}{3pt}
\begin{table*}[ht]
		\centering
		\caption{Spectral analysis of three pairs of surface brightness edges.}
		\renewcommand{\arraystretch}{1.5}
		\label{tab:fronts}
		\begin{tabular}{cccc|ccccc}
			\hline
			Shock & Ellipticity (P.A.) & $\theta_{1} - \theta_{2}$  & $\chi^{2}/D.o.f.$ & Side & R$_{\text{i}}$ & R$_{\text{o}}$  & $kT$ &    $p_{\text{ICM}}$ \\
			& & & & & [kpc ($''$)] & [kpc ($''$)] & [keV] &   [10$^{-10}$ erg cm$^{-3}$]\\
			\hline

		 \multirow{2}{*}{S1 (53.9 kpc)} & \multirow{2}{*}{1.2 (345$^{\circ}$)} & \multirow{2}{*}{130$^{\circ}$ - 190$^{\circ}$} & \multirow{2}{*}{0.97} & \textit{downstream}& 45.6 (9.3) & 53.9 (11) & 8.9$^{+0.9}_{-0.8}$   &  12.9$^{+1.2}_{-1.1}$ \\

		& & & & \textit{upstream} & 53.9 (11) & 63.7 (13)  & 6.6$^{+0.7}_{-0.7}$  & 7.1$^{+0.7}_{-0.7}$ \\
		
		\hline
		
		\multirow{2}{*}{S2 (50.5 kpc)} & \multirow{2}{*}{1.2 (345$^{\circ}$)} & \multirow{2}{*}{310$^{\circ}$ - 15$^{\circ}$} & \multirow{2}{*}{1.08} & \textit{downstream}& 40.2 (8.2) & 50.5 (10.3) & 5.4$^{+0.4}_{-0.3}$   &  9.8$^{+1.0}_{-0.9}$ \\

		& & & & \textit{upstream} & 50.5 (10.3) & 68.6 (14)  & 4.8$^{+0.2}_{-0.2}$  & 6.5$^{+0.6}_{-0.6}$ \\
		
		\hline
		
		\multirow{2}{*}{S$_{\text{in}}$ (52.4 kpc)} & \multirow{2}{*}{1.2 (345$^{\circ}$)} & \multirow{2}{*}{0$^{\circ}$ - 360$^{\circ}$} & \multirow{2}{*}{0.98} & \textit{downstream}& 44.1 (9) & 52.4 (10.7) & 5.9$^{+0.4}_{-0.4}$   &  10.1$^{+0.7}_{-0.7}$ \\

		& & & & \textit{upstream} & 52.4 (10.3) & 68.6 (14)  & 5.0$^{+0.2}_{-0.2}$  & 5.4$^{+0.3}_{-0.3}$ \\
		


		\hline

		\multirow{2}{*}{S3 (79.4 kpc)} & \multirow{2}{*}{1.1 (20$^{\circ}$)} & \multirow{2}{*}{200$^{\circ}$ - 270$^{\circ}$} & \multirow{2}{*}{0.97} & \textit{downstream}& 63 (12.8) & 79.4 (16.2) & 7.9$^{+0.8}_{-0.8}$   &  6.3$^{+0.7}_{-0.7}$ \\

		& & & & \textit{upstream} & 79.4 (16.2) & 108 (22)  & 6.4$^{+0.3}_{-0.4}$  & 3.8$^{+0.2}_{-0.3}$ \\

		\hline

		\multirow{2}{*}{S4 (81.8 kpc)} & \multirow{2}{*}{1.1 (20$^{\circ}$)} & \multirow{2}{*}{45$^{\circ}$ - 110$^{\circ}$} & \multirow{2}{*}{0.91} & \textit{downstream}& 63.7 (13) & 81.8 (16.7) & 6.7$^{+0.6}_{-0.6}$   &  6.3$^{+0.6}_{-0.6}$ \\

		& & & & \textit{upstream} & 81.8 (16.7) & 118 (24)  & 5.4$^{+0.4}_{-0.4}$  & 2.7$^{+0.2}_{-0.2}$ \\

		\hline

		\multirow{2}{*}{S5 (136 kpc)} & \multirow{2}{*}{1.1 (102$^{\circ}$)} & \multirow{2}{*}{140$^{\circ}$ - 252$^{\circ}$} & \multirow{2}{*}{1.03} & \textit{downstream}& 113 (23) & 149 (30.5) & 9.5$^{+1.7}_{-1.6}$   &  2.7$^{+0.4}_{-0.4}$ \\

		& & & & \textit{upstream} & 149 (30.5) & 226 (46.2)  & 6.9$^{+0.8}_{-0.8}$  & 1.4$^{+0.2}_{-0.1}$ \\

		\hline

		\multirow{2}{*}{S6 (132 kpc)} & \multirow{2}{*}{1.1 (102$^{\circ}$)} & \multirow{2}{*}{304$^{\circ}$ - 87$^{\circ}$} & \multirow{2}{*}{0.99} & \textit{downstream}& 102 (21) & 142 (28.9) & 8.3$^{+1.7}_{-1.2}$   &  2.9$^{+0.8}_{-0.5}$ \\

		& & & & \textit{upstream} & 142 (28.9) & 216 (44)  & 5.9$^{+0.7}_{-0.7}$  & 1.3$^{+0.2}_{-0.2}$ \\

			\hline
		\end{tabular}
		\tablecomments{(1) Name of the edge and distance from the center (measured from the front mid-aperture); (2) ellipticity and position angle of the sector used as spectral extraction region; (3) angular range intersecting the position of the edge; (4) $\chi^{2}/d.o.f.$; (5) side of the edge (within - downstream, outside - upstream); (6-7) inner and outer radius of the annular sector (along the major axis of the ellipse); (8) deprojected ICM temperature; (9) deprojected ICM pressure.}
\end{table*}	

\end{appendix}

\end{document}